\documentclass[apj, iop]{emulateapj}
\usepackage{amsmath}
\usepackage{amssymb}
\usepackage{graphicx,graphics}
\usepackage[english]{babel}
\usepackage{color}

\def\PGPU{$\varphi$GPU }
\def\PGRAPE{$\varphi$GRAPE }
\begin{document}

\title{SMBH in Galactic Nuclei with Tidal Disruption of Stars}

\author{
       Shiyan Zhong   \altaffilmark{1},
       Peter Berczik  \altaffilmark{1,2,3},
       Rainer Spurzem \altaffilmark{1,2,4}
       }

\altaffiltext{1}{National Astronomical Observatories of China, Chinese Academy of Sciences, 20A Datun Rd., Chaoyang District, 100012, Beijing, China}
\altaffiltext{2}{Astronomisches Rechen-Institut, Zentrum f\"ur Astronomie, University of Heidelberg, M\"onchhofstrasse 12-14, 69120, Heidelberg, Germany}
\altaffiltext{3}{Main Astronomical Observatory, National Academy of Sciences of Ukraine, 27 Akademika Zabolotnoho St., 03680, Kyiv, Ukraine}
\altaffiltext{4}{Kavli Institute for Astronomy and Astrophysics, Peking University, Beijing, China}

\shorttitle{Tidal Disruption by SMBH}
\shortauthors{Zhong, Berczik, Spurzem}

%%%%%%%%%%%%%%%%%%%%%%%%%%%%%%%%%%%%%%%%%%      ABSTRACT

\begin{abstract}

Tidal Disruption of stars by super massive central black holes from dense star clusters is modeled by high-accuracy direct $N$-body simulation. The time evolution of the stellar tidal disruption rate, the effect of tidal disruption on the stellar density profile and for the first time the detailed origin of tidally disrupted stars are carefully examined and compared with classic papers in the field. Up to 128k particles are used in simulation to model the star cluster around the super massive black hole, we use the particle number and the tidal radius of black hole as free parameters for a scaling analysis. The transition from full to empty loss-cone is analyzed in our data, the tidal disruption rate scales with the particle number $N$ in the expected way for both cases. For the first time in numerical simulations (under certain conditions) we can support the concept of a critical radius of ~\citet{FR1976}, which claims that most stars are tidally accreted on highly eccentric orbits originating from regions far outside the tidal radius. Due to the consumption of stars moving on radial orbits, a velocity anisotropy is founded inside the cluster. Finally we make an estimation for the real galactic center based on our simulation results and the scaling analysis.

\end{abstract}

%%%%%%%%%%%%%%%%%%%%%%%%%%%%%%%%%%%%%%%%%%


%%%%%%%%%%%%%%%%%%%%%%%%%%%%%%%%%%%%%%%%%%

\section{Introduction}

Most galaxies show evidence for a
super massive black hole (SMBH) residing in their center (cf. e.g.
~\citet{GH2009}). All detections for SMBH are indirect, with the strongest
case in the
center of our own galaxy ~\citep{GEG2010}. Since the detection of quasars in the
60's the huge energy output of active galactic nuclei (AGN) has fascinated
astronomers; the idea that it is powered by a central SMBH which accretes gas
originates already from the same time ~\citep{C1967,S1970,R1984}. After more
than half a century's development, SMBH masses are measured using
different methods in the local universe as well as in cosmological
distances~\citep{GH2009}. Quasars and AGN have faded in the local
universe, and there are many galaxies in our local environment, which have SMBH,
but are not active ~\citep{KB2009,FCD2006,T2002,GBB2000}. These and other papers
have established correlations between the central SMBH properties and those of their host
galaxies, such as e.g. the $M_{\bullet}- {\sigma_{*}}$ relation. The relations
suffer from strong scatter which can be either an observational error or a physical
variation. Some of the best measurements are
direct kinematical measurements of individual stars, as in our Milky Way,
or e.g. water maser measurements such as of e.g. NGC4258 ~\citep{MMH1995,NM1995,GHB1995}.
In most cases determinations of black hole masses and stellar velocities are not as
accurate, though.

Non-active galaxies usually do not emit any electromagnetic
radiation originating directly from the vicinity of their SMBH's Schwarzschild radius.
However, sometimes the stellar system surrounding the SMBH delivers a star close
enough to the black hole, that strong tidal forces will
disrupt the star. The gaseous debris will be heated by dissipation and
release a sudden burst of thermal radiation, typically in the X-ray region;
its energy
originates from the deep gravitational well of the SMBH.

Tidal disruption (TD) of stars has been proposed already about
50 years ago ~\citep{H1975,H1976,FR1976}. These events occur at a
distance very close to black hole, so they can help us constrain the black
hole's mass and other parameters like its spin more accurately. There are
already some candidates of tidal disruption events
~\citep{Wang2012,KM2008,Komossa2002}. More signatures of tidal disruption events
may still be hidden in data archives.

Since tidal disruptions per galaxy are rare, and many events very distant
~\citep{Komossa2002}, a thorough theoretical study is important to derive
reliable predictions for event rates. We need to know what is the expected tidal
disruption rate as a function of black hole mass, stellar populations and host
galaxy morphology, and also a better understanding of individual tidal
disruption events. ~\citet{G_RR2012} show that usually only some fraction of the
gas from the disrupted star is accreted by the SMBH, while another part gets
large enough energy to escape. The escaping gas from tidal disruption could be
detected in other wavebands e.g. by radio astronomy (LOFAR,
~\cite{VFF2010,VKF2011}).
Like most of previous authors in the field,
however, we assume in this work just a 100\% accretion of stellar debris to the
SMBH.

Recent research has focused on the cosmological growth of SMBH in galactic
nuclei. There is increasing evidence that galaxy mergers in the very young
universe, based on the merger statistics of a standard $\Lambda$-CDM (cold dark
matter) model, provide a mechanism to feed large amounts of gas to the central
SMBH in these galaxies. The general conclusion is that tidal disruption of stars
only plays a minor role for growth of the most massive black holes, and that the
observed correlations between the central black hole mass of galaxies and
parameters of their surrounding galaxies (velocity dispersion, bulge mass, mass
deficit) can be reproduced in such a scenario ~\citep{HKB2010, JBN2009}. Similar
conclusions have already been reached much earlier by simpler or semi-analytic models
~\citep{HR1993,YT2002}. This lends support to the idea of the
``anti-hierarchical'' growth of black holes (the distribution of most
luminous and massive active galactic nuclei peaks at higher redshifts
~\citep{HSN2012}). The current spatial and time resolution of simulations,
while excellent on the scales of galactic mergers (of order parsec in some
cases), is still far from any required resolution to resolve mass, energy and
angular momentum transport deep down towards the black hole's Schwarzschild
radius (of order $10^{-6}$ pc for typical SMBH of the mass scale like in our own
galaxy) or stellar tidal disruption radius (of order $10^{-5}$ pc for solar type
stars). The simulations also do not even resolve the gravitational influence spheres
of the SMBH usually.

But it is possible that for lower mass SMBH (about $10^6 ~{\rm M}_\odot$) tidal
disruption of stars is a major constituent of their growth, because many of
these galaxies are quiescent and show no trace of large scale gas accretion; our
own galaxy is a good example for such an extremely gas poor object
~\citep{NCS2007,CNM2008}. \citet{MMH2006} even argue that most of the X-ray
luminosity of low-luminosity AGN may stem from tidal disruptions.

The theoretical study of black hole mass growth due to tidal disruption has a
history of many decades. ~\citet{BW1976,BW1977} and \citet{DO1977a,DO1977b} were
the pioneers to use tidal disruption as inner boundary condition near the SMBH
to determine the steady state density distribution of stars around it. This has
later become known as a ``Bahcall-Wolf'' cusp, with a density power-law exponent
of -7/4 (for equal mass stars). It is
less known in the astrophysical community that such a solution was already known
in plasma
physics for the distribution of electrons around a central massive nuclear particle
of the opposite charge in the center \citep{G1964},
only cited in \citet{DO1977a}. At about the same time also ~\citet{FR1976,
SL1976}, and \citet{LS1977} discussed this solution using some thermodynamical
arguments by equating a proper timescale of energy transport in the cusp with
the stellar dynamical two-body relaxation time scale. Such models, later
completed to model an entire star cluster, not only the cusp in the vicinity of
the SMBH, became known as gaseous models of star clusters
\citep{LS1991,ST1995,AFS2004,SAS2011}. Direct solutions of Fokker-Planck models
have also been extensively used to study the problem \citep{CK1978,MCD1991}, as
well as Monte Carlo methods
\citep{SM1978,MS1979,MS1980,DS1983,Umbreit2012,GUB2012,UR2012}. All methods
converge to similar results, which is not very surprising, since all of them are
based on an implementation of loss-cone theory as originally given by
\citet{FR1976, BW1976} and \citet{CK1978}, see also next section.

Standard loss-cone theory as in the cited papers deals
with refilling the stellar orbits
which intersect the tidal radius at which tidal forces of the SMBH
will disrupt them. In spherically symmetric systems there is only two body relaxation,
which can refill these orbits by diffusion of angular momentum of stars.
However, in case of deviations from
spherical symmetry time scales of angular momentum
diffusion may become dramatically faster than by two-body relaxation alone.
This had been realized early on by \citet{NS1983} and
\citet{Malkov1993}, and confirmed by a combination of Schwarzschild's method,
some special direct $N$-body simulations and analytic or semi-analytic reasoning
\citep{PM2002,PM2004,MP2004}, and studied further by \citet{WM2004} and
\citet{MW2005} for the case of the loss-cone around a supermassive binary black
hole. \citet{Berczik2006,KBB2012,KHBJ2013} confirmed that the so-called 'final parsec problem' of
shrinking binary black holes in galactic nuclei until they can coalesce
relativistically, could already be solved in just axisymmetric galactic nuclei.

Except for some pioneering but preliminary results by \cite{MW2005} the question
how tidal disruption rates (and their observational counterparts, X-ray flares)
might be affected by the presence of a binary black hole has only
recently been studied, by some semi-analytic modelling, \citep{Chen2009,Chen2011,Liu2009,Chen2008},
by direct numerical solutions of a 2D Fokker-
Planck equation and comparisons with direct $N$-body models
\citep{Fiestas2010,Fiestas2012}. Finally the impact of recoils of
SMBH from galactic nuclei \citep{Li2012,GM2009} and the influence of a presence
of a central gaseous disk around the SMBH \citep{JYM2012} on the tidal
disruption rates have been studied.

In our work here we focus on a high resolution direct $N$-body model of the
problem. It has the smallest number of inherent physical approximations and
allows to test and measure in detail processes connected to the loss-cone
accretion of stars to the SMBH and transport processes of mass and energy in the
central stellar cusp, along with the self-consistent evolution of the cusp, the
black hole growth and feedback of tidal disruption to the stellar density
profile. Such models have been rather difficult in the past, because they
require large particle numbers (say a million) in order to correctly resolve
two-body relaxation around a few million solar mass central black hole,
including all regions from where loss-cone stars may come from (stars bound and
unbound to the SMBH).
\citet{BME2004a,BME2004b,BHPM2006} have used GRAPE hardware
\citep{Sugimoto1990,Makino1997GRAPE-4} for some pioneering studies,
but in spite of using the best available hardware at the time for
the direct $N$-body problem their numerical
resolution was not quite sufficient yet (maximum order $10^5$ stars),
and not a large number of statistically independent cases could be simulated. Using
NBODY6 with a single GPU accelerator helped to increase the particle number a
little (half a million) and simulate a larger number of cases (\citet{BBK2011},
henceforth BBK).

Our study uses the massively parallel $N$-body code \PGPU ~\citet{BNZ2011}, see
for more detailed discussions also ~\citet{SBZ2012, KBB2012} and
~\citet{HGM2007} in conjunction with many GPU's in parallel on a special
supercomputer in China. As in BBK we will present in this paper a detailed
parameter and scaling study of the problem of tidal disruption of stars near
SMBH in conjunction with a fully self-consistent simulation of the surrounding
stellar cluster and the black hole motion. Our particular interest here, which
goes beyond BBK, is in the origin of stars tidally accreted by the central SMBH,
which is a key element to verify the classic loss cone theories.

First we will give a brief introduction to loss-cone theory. Then we describe
our model setups, as a first step we use Plummer model to modeling the galactic
center (Section 2). The third part (Section 3) is analyzing of the result data.
Using our $N$-body simulation data, we focused on following issues: TDR,
stellar spatial distribution and velocity dispersion evolution, orbital
parameters of the disrupted stars. In Section 4, we will make an estimation of
TDR in real world based on our results. And final conclusions are given in
Section 5.

%%%%%%%%%%%%%%%%%%%%%%%%%%%%%%%%%%%%%%%%%%

\section{Loss-cone theory}

Stars are inside the loss cone, if the peribothron of their orbit (pericenter
distance to the supermassive central black hole, SMBH) is close enough to the
SMBH so that tidal forces disrupt the star. The key question is how many loss
cone stars there are and how their population is changing.
Before loss-cone theory,
~\citet{P1972} has already presented a scenario based on energy
arguments. In these arguments, stars which have an energy smaller than $E_t
\equiv - GM_{\rm bh}/{r_t}$ will be disrupted by the black hole. It was assumed
that the stars are on an approximately circular orbits. Stars lose energy near
the SMBH through two-body interactions and sink towards the black hole. Here the
tidal radius $r_t$ is given as

\begin{equation}
r_t = \alpha r_\star \frac{3}{5-n} \Bigl(\frac{M_{\rm bh}}{ m_\star} \Bigr)^{\frac{1}{3}}
\end{equation}

where $r_\star$ and $m_\star$ are the radius and mass of the star approaching
the black hole, $n$ is the polytropic index of the internal stellar structure
\citep{H1975, H1976}. $\alpha$ is a free parameter we use for our scaling study.
Later it was found that angular momentum diffusion of stars further away from the
tidal radius is a much
faster process for replenishment of loss cone orbits \citep{FR1976,LS1977,BW1976}.
A star with relatively high energy
(not strongly bound to the black hole), but very small angular momentum, has
a nearly radial orbit - within a dynamical time it can get to a pericenter smaller than $r_t$, where
it will be tidally disrupted. So,
the region from which tidally disrupted stars can come from is much larger than just the vicinity of
the black hole, and the
number of stars which can be potentially disrupted could be larger than originally expected.

Energy and angular momentum conservation for a stellar orbit in a spherically
symmetric potential leads to the following expression for the maximum tangential
velocity of a star in the loss cone:

\begin{equation}
|v_t| < v_{\rm lc}(r)=\frac {r_t}{\sqrt {r^2-r_t^2}}
\cdot \sqrt {2[ \phi (r_{t}) - \phi (r)] + v_{r}(r)^2}
\label{eq.v_lc}
\end{equation}

\citep{AFS2004}. We assume that stars in the loss cone will be tidally disrupted in one
dynamical time, and their mass is added to the SMBH mass. Actually the process of stellar disruption and accretion of the gas
to the SMBH is much more complicated in reality (see e.g. \citet{G_RR2012}, and
further references therein), but like other authors in this field we follow
the simple model,
that once a star passes inside $r_t$ its mass will be
immediately added to the black hole.

An opening angle of the loss cone can be defined by $\theta_{\rm lc} =
v_{\rm lc}(r)/\sigma_r$, where $\sigma_r$ is the radial velocity dispersion of
the stars in the system at radius $r$. After one orbital time the loss cone will
be empty, and the future tidal disruption rate will depend on the rate with
which angular momentum diffusion of stars is repopulating the loss cone.
In \citet{FR1976}, they defined a
diffusion angle $\theta_{\rm D}$ per dynamical time, to measure the efficiency of
repopulating the loss cone by two-body relaxation:

\begin{equation}
\theta_{\rm D}^{2} = \frac{t_{\rm dyn}}{t_{\rm relax}}
\label{theta_D}
\end{equation}

\noindent

In regions where $\theta_{\rm lc}>\theta_{\rm D}$, repopulating is inefficient.
The typical time for repopulating is about $R^2 t_{dyn}$ where $R=\theta_{\rm
lc}/\theta_{\rm D} $. In this case we have an empty loss-cone. In regions where
$\theta_{\rm lc}<\theta_{\rm D}$, a star can enter and go out of loss-cone
freely within a crossing time without being disrupted, the loss-cone concept
loses its significance (this regime is sometimes called the pinhole regime). So
we expect that only a few tidally disrupted stars are coming from the above two
regions and most of them should come from the regions where $\theta_{\rm lc}$
roughly equals to $\theta_{\rm D}$. And we use the definition of critical radius
$r_{crit}$ proposed by \citet{FR1976} which reads $\theta_{\rm lc}=\theta_{\rm
D}$. The reader interested in more details about this model is also referred to
\citet{AFS2004}, who give a very detailed description.

%%%%%%%%%%%%%%%%%%%%%%%%%%%%%%%%%%%%%%%%%%

%%%%%%%%%%%%%%%%%%%%%%%%%%%%%%%%%%%%%%%%%%
\section{Model Description}
%%%%%%%%%%%%%%%%%%%%%%%%%%%%%%%%%%%%%%%%%%

We adopt the unit definition from \citet{HM1986}, namely $G$ = $M$ = 1 and $E$ = $-1/4$, where $G$ is the gravitational constant, $M$ is the total mass of the model cluster and $E$ is the total energy. In our $N$-body models we assume all the particles have the same mass, e.g. $m=1/N$, here $N$ is the total particle number. We choose different values of $N$ to investigate the N-dependence of our simulations. Currently adopted $N$ is from 16K to 128K (see Table~\ref{tablemod}, please note through out this paper we define $1K = 1024$ due to technical reason). Initially the stars are distributed following a Plummer model for simplicity, which is generated using the method given in ~\citet{AHW1974}. One of the central particles is regarded as the central black hole, we change its initial mass to 0.01 (in our scaled units) and put it to rest at the origin (zero velocity). This black hole particle can move freely and has a finite tidal radius $r_{t}$. Once a star comes into the tidal radius, it will be removed from the stellar system, by the meantime the black hole gain that star's mass and linear momentum. The initially flat core of the Plummer model will adjust in a few time units to the central black hole's gravity. All our stars have equal mass and are single (no binaries). Our current code is not able to deal efficiently with strongly bound binaries; star clusters near SMBH have a very high velocity dispersion, so not many binaries survive (but see e.g. ~\citet{LYL2007} for the importance of some binaries in the galactic center for ejection of hypervelocity stars).

For simplicity, we adopt three fixed tidal radius: $10^{-3}$, $5\times10^{-4}$ and $10^{-4}$. Since the BH mass changes within one order of magnitude during the simulation, relative changes in tidal radius is tiny (notice that $r_{t}\propto(M_{\bullet})^{1/3}$), so fixed tidal radius can be a safe approximation. The tidal radius we used here are larger than the actual value, typically boosted by a factor of $10^{3}$-$10^{4}$. The reasons for doing so are 1) our softening parameter $\epsilon$ is $10^{-5}$, $r_{t}$ must larger than $\epsilon$; 2) we need sufficient TD events so that we can do statistical research. We will see in the rest of this paper that these 3 choices of tidal radius already give us different results.

%%%%%%%%%%%%%%%%%%%%%%%%%%%%%%%%%%%%%%%%%%%%%%%%%%%%
\begin{table}[htbp]
\begin{center}
\caption{Full set of our model runs.\label{tablemod}}
\begin{tabular}{c|ccc}
  \tableline
  Model &   N/K     &   $r_{t}$       &   T      \\
  \hline
  N00   &   16      &  $10^{-3}$       &   500    \\
  N01   &   16      &  $10^{-4}$       &   500    \\
  N02   &   16      &  $5\times10^{-4}$&   500    \\
  N10   &   32      &  $10^{-3}$       &  1000    \\
  N11   &   32      &  $10^{-4}$       &  1000    \\
  N12   &   32      &  $5\times10^{-4}$&  1000    \\
  N20   &   64      &  $10^{-3}$       &  1000    \\
  N21   &   64      &  $10^{-4}$       &  1000    \\
  N22   &   64      &  $5\times10^{-4}$&  1000    \\
  N30   &  128      &  $10^{-3}$       &  2200    \\
  N31   &  128      &  $10^{-4}$       &  2500    \\
  N32   &  128      &  $5\times10^{-4}$&  1000    \\
  \tableline
\end{tabular}
\tablecomments{First column: Model series number. Column 2 : Particle number in
the unit of K(=1024). Column 3: black hole's tidal radius. Column 4 : total
integration time. $r_{t}$ and $T$ are in model unit.}
\end{center}
\end{table}
%%%%%%%%%%%%%%%%%%%%%%%%%%%%%%%%%%%%%%%%%%%%%%%%%%%%

Simulations are running for at least 1 half-mass relaxation time ($t_{rh}$), which
can be estimated as follow:

\begin{equation}
%t_{rh} = 0.065 \frac{\sigma^{3}}{{\mathrm{G}}^{2} m \rho \ln(\Lambda)}
t_{rh} = 0.1 \frac{N}{\ln(\Lambda N)} t_{dyn}
\label{trelax}
\end{equation}

\noindent where $\Lambda=0.11$ \citep{GS1994} and $t_{dyn}$ is the dynamical
time scale at half-mass radius. Some of the models are running up to 2 $t_{rh}$
to see their long term evolution. Table~\ref{tabletrh} shows the approximate
value of $t_{rh}$ for models with different particle numbers.

Our simulations focus on the innermost parts of a galactic nucleus. Scale is less
than some 10-100 times the gravitational influence radius $r_h$ (see for
a definition Sect.~\ref{Scaling}) of the central SMBH; therefore there is no need
to take any dark matter into account. We actually are interested in this paper
only in the stellar dynamical processes in this region, so there is no interstellar
gas or clouds in our models for the current time.

All simulations are using the \PGRAPE code \citep{BNZ2011}, which runs with high
performance (up to 350 Gflop/s per GPU) on our GPU clusters in Beijing (NAOC/CAS)
and Heidelberg (ARI/ZAH) \citep{SBZ2012, Berczik2013}.
The code is a direct $N$-body simulation package, with a high order
Hermite integration scheme and individual block time steps. A direct $N$-body code
evaluates in principle all pairwise forces between the gravitating particles,
and its computational complexity scales asymptotically with $N^2$; however, it
is {\em not} to be confused with a simple brute force shared time step code, due
to the block time steps. We refer more interested readers to a general discussion
about $N$-body codes and their implementation in \cite{spurzem2011a,spurzem2011b}. The
present code is well tested and already used to obtain important results in our
earlier large scale few million body simulation ~\citep{KBB2012}.

Our code allows for an initial dynamical relaxation of the system, before switching
on the tidal disruption routine (hereafter TD routine). It turns out that for the
results presented here the initial relaxation have tiny effect on TDR (we tested 1/4
of $t_{rh}$ as time interval allowed for initial relaxation). Therefore we did not
use the initial relaxation mechanism for all models of this paper.

%%%%%%%%%%%%%%%%%%%%%%%%%%%%%%%
\begin{table}[htbp]
\begin{center}
\caption{$t_{rh}$ for different set of models.\label{tabletrh}}
\begin{tabular}{c|c}
  \tableline
  N/K     &   $t_{rh}$      \\
  \hline
  16       &   164    \\
  32       &   300    \\
  64       &   553    \\
  128      &  1025    \\
  \tableline
\end{tabular}
\tablecomments{Column 1: particle number in the unit of K(=1024).
Column 2: approximate value of initial half-mass relaxation time
in model unit.}
\end{center}
\end{table}
%%%%%%%%%%%%%%%%%%%%%%%%%%%%%%%

%%%%%%%%%%%%%%%%%%%%%%%%%%%%%%%
\section{Results \& Discussion}

\subsection{Tidal Disruption Rate}
\label{TDR}
%%%%%%%%%%%%%%%%%%%%%%%%%%%%%%%

Key questions about tidal disruption (henceforth TD) are the rate of accretion
of mass, energy and angular momentum to the SMBH with time. In this paper we
focus on the mass and number accretion rate by TD, and how does it vary with
the particle number and the tidal radius.

\begin{figure}[htbp]
  \begin{center}
  \includegraphics[width=0.8\columnwidth]{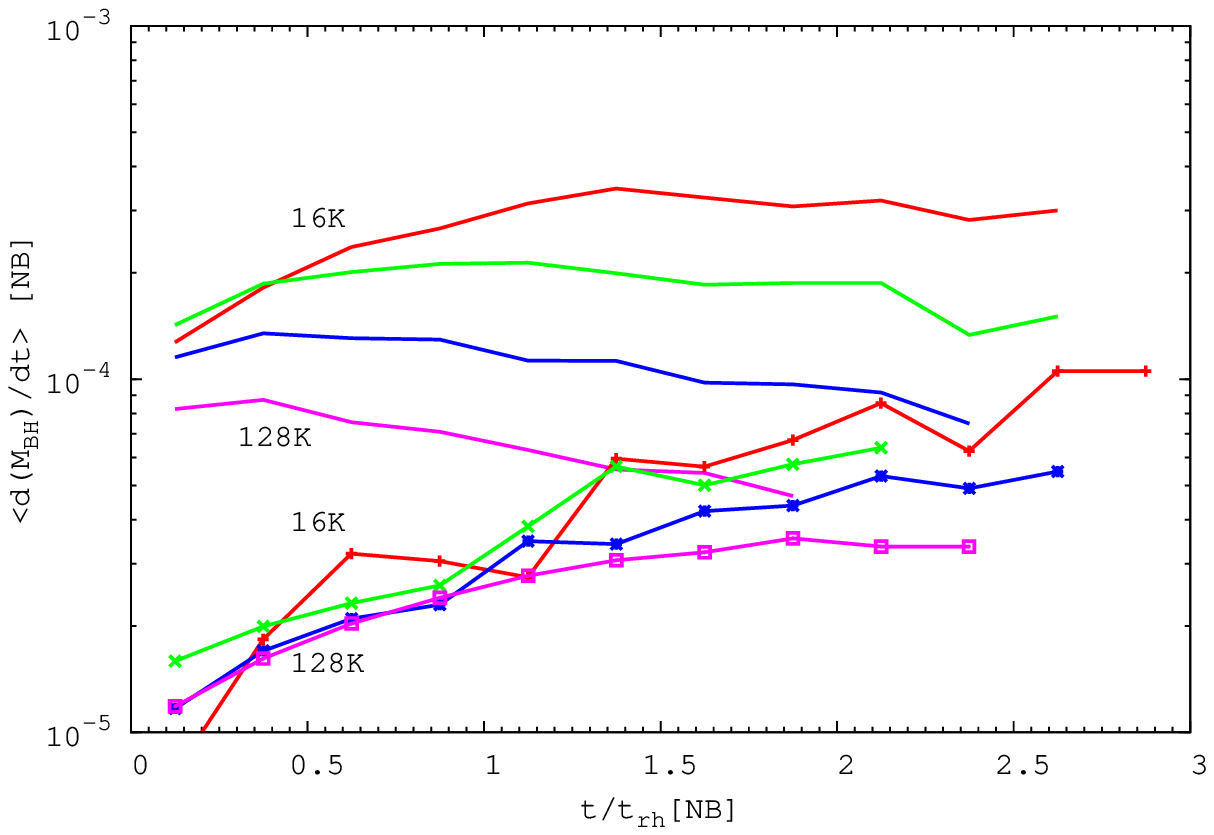}
  \includegraphics[width=0.8\columnwidth]{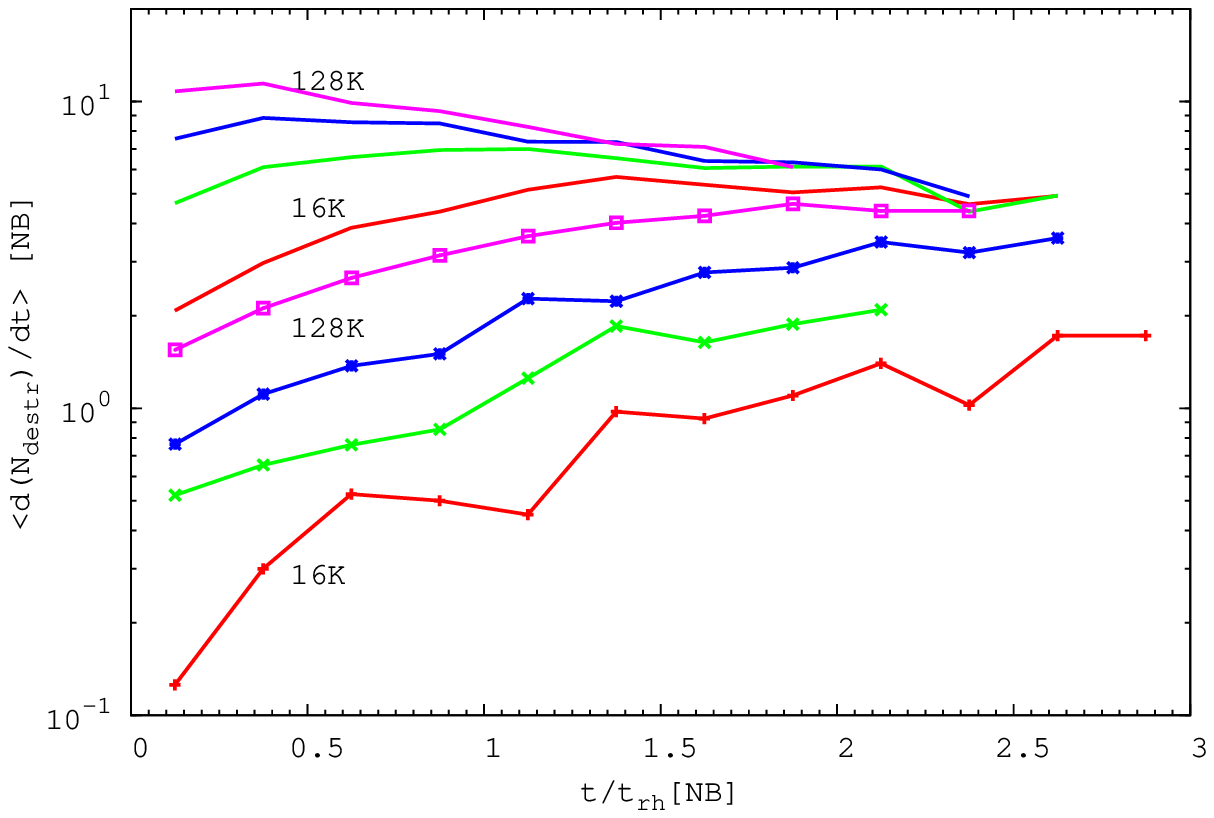}
  \end{center}
  \caption{Disruption rate of different models. $x$ axis is time expressed in unit of initial half-mass relaxation time ($t_{rh}$), $y$ axis is the averaged disruption rate in a given time range (i.e. 1/4 $t_{rh}$). In upper panel the unit for disruption rate is mass increase per unit time, while in lower panel it is number of disrupted stars per unit time. TD curves with symbols stand for $r_{t}=10^{-4}$. TD curves without symbols are for $r_t=10^{-3}$. }
  \label{fig_Mdot}
\end{figure}

Figure~\ref{fig_Mdot} shows the tidal disruption rate as a function of time, both
in terms of mass and particle number accretion rate. The
$x$ axis is time expressed in unit of initial half-mass relaxation
time $t_{rh}$, $y$ axis is the averaged disruption rate in a given time
interval (i.e. 1/4 $t_{rh}$). From the lower panel of Figure~\ref{fig_Mdot}
one can see how the does TDR depend on tidal radius and particle number.

The TDR curves can be divided into 2 groups, depending on their tidal radius $r_{t}$. In each group, the 4 curves are initially well separated, showing some N-dependence of TDR. At early stage, all curves are rising due to the cusp formation until they reach a turning point. Systems with larger $r_{t}$ (i.e. $10^{-3}$ ) reach the turning point earlier. And models with larger particle number reach turning point earlier. Afterwards the TDR curves converge and begin to decline, in this stage TDR only weakly depend on $N$. In the case of $r_{t}=10^{-4}$, the trend of convergence is less pronounced, but we find that if integrated for longer time these 4 curves will show same behavior as observed in $r_{t}=10^{-3}$ models.

Why do TDR curves behave like this? We can get some hints from the loss-cone theory. At the beginning of simulation, we have an isotropic velocity distribution through out the whole cluster so the loss-cone is full everywhere. As the simulation goes on, stars inside the loss-cone are disrupted quickly by the black hole and loss-cone becomes empty. The typical time scale for this process is one crossing time -- but repopulating the loss-cone requires much longer time, i.g. relaxation time scale. According to this, we expect the TD curves to drop from the very beginning. However we see in the simulation the TDR curves keep rising for at least 0.5$t_{rh}$. Does loss-cone theory fails in our case? The classical loss cone theory assumes that the black hole is fixed at the origin; in contrast to this, our simulations allow the black hole to move freely. We think that it is the black hole's motion caused the deviation from standard loss-cone theory in our models.

\begin{figure}[htbp]
  \begin{center}
  \includegraphics[width=0.8\columnwidth]{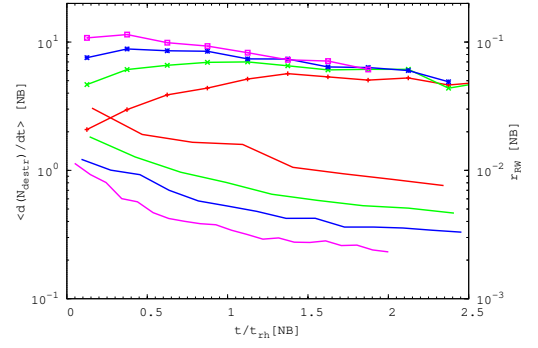}
  \end{center}
  \caption{$x$ axis is time expressed in unit of initial half-mass
    relaxation time ($t_{rh}$), left $y$ axis is the averaged
    disruption rate in a given time range (i.e. 1/4 $t_{rh}$). The
    unit for disruption rate is number of disrupted stars per unit
    time. Right $y$ axis is the Brownian motion amplitude. Curves with
    symbols are TDR for $r_{t}=10^{-3}$, from bottom to top,
    respond for $N$ = 16, 32, 64 and 128K. Curves without symbols shows
    black hole's Brownian motion amplitude for $N$ = 16, 32, 64 and 128K (from top to bottom)}
  \label{fig_Ndot_bhxv}
\end{figure}

Figure~\ref{fig_Ndot_bhxv} gives a good example to see how the black hole's motion affects our results. At the beginning SMBHs in all 4 models have large motions. This motion can be described as Brownian motion ~\citep{LT1980}. Due to this motion the SMBH will always ``see'' a full loss-cone which is not efficiently repopulated by two-body relaxation but by its vigorous movement.

When loss-cone is full, we follow the description in Merritt (2001), but note the early discussion of Brownian motion of SMBH in ~\citet{LT1980}; they consider an SMBH binary, but there is an analogous argument for single SMBH's. The flow of stars which make an encounter with the SMBH is estimated with a standard ansatz as

\begin{equation}
\dot N \propto n \Sigma v = \frac{\rho}{m} \Sigma v = N \rho \Sigma v
\label{Ndot_full_lc}
\end{equation}

where $\rho$ and $m$ are the local stellar mass density and the average stellar mass in the vicinity of the SMBH, and $\Sigma$, $v$ the cross section of the SMBH-star interaction ($\Sigma = \pi r_t^2$ to first order, without gravitational focusing) and the relative velocity between stars and the SMBH. If we increase the particle number by keeping the total mass constant, the quantities $\rho$, $\Sigma$ and $v$ will remain constant, so we expect the particle number accretion rate to scale as $\dot N \propto N$, as is observed initially in our simulations. Note that this approximation is only valid if there are always enough stars for the SMBH to interact with - if the loss cone is getting empty another approximation is needed. The moving SMBH can also generate additional perturbations on the surrounding stars' energy and angular momentum, change their distribution in phase space and configuration space, which will cause deviations from our simple model in the numerical simulation.

As time goes by, black hole gains mass from the disrupted stars and as a consequence its Brownian motion amplitude decreases. Once the amplitude drops below a threshold (in the case of $r_t=10^{-3}$ roughly 0.01, one order of magnitude larger than $r_t$, this empirical relation might also be true for other $r_t$), we can treat the BH as a static object and the system quickly enters the classic empty loss-cone stage. In this stage TDR becomes weakly depend on $N$, since

\begin{equation}
\dot N \propto \frac{N}{t_{\rm relax}}
     \propto \frac{N}{N/\ln( \Lambda N)}
     = \ln( \Lambda N)
\label{Ndot_empty_lc}
\end{equation}

Note that in Fig.~\ref{fig_Mdot} we show both the mass and the particle number accretion rates $\dot M$ and $\dot N$; since ${\dot M} = m {\dot N} = {\dot N} / N$ their scaling behavior with respect to $N$ is just inverted - due to the full loss-cone we have initially ${\dot M} = {\rm const.}$, and later on ${\dot M} \propto 1/N$ for the empty-loss cone. From Fig.~\ref{fig_Ndot_bhxv} we see that the time when SMBH's motion becomes sufficiently small is closely related to the mass ratio between stars and the SMBH ($\gamma := m/M_{\bullet}$). In model N30 ($N$ = 128K, $\gamma = 7.62\cdot 10^{-4}$ ), TDR curve is flattening right from the beginning, which is a signal of depleting loss-cone. On the meantime, one can see the Brownian motion in N30 model already falls below the threshold. This is no surprise since in this model we have a small mass ratio between stars and SMBH from the beginning. So it will not take long time to bring the BH to ``rest''. While in the model N00 ($N$ = 16K, $\gamma = 6.10\cdot 10^{-3}$ ), the BH spend much more time to get to ``rest'' because of the relatively large mass ratio of stars to the SMBH initially. We see in Fig.~\ref{fig_Mdot} that each of the four curves of different $N$ but otherwise same parameters converge at the time when Brownian motion damps below the threshold, and the TDR also begins to drop. We interpret this secular drop of the TDR as a consequence of changing parameters of the star cluster (density, velocity dispersion) at the origin of most of the tidally disrupted stars (the critical radius, see below).

\begin{figure}[htbp]
  \begin{center}
  \includegraphics[width=\columnwidth]{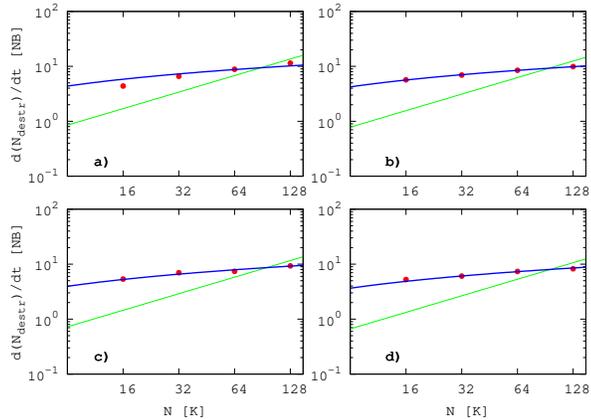}
  \end{center}
  \caption{Particle number dependence of TDR. $x$ axis is particle numbers in unit of K. $y$ axis is the averaged disruption rate in a given time range (i.e. 1/4 $t_{rh}$). Tidal radius is $10^{-3}$. TDR data points in these 4 panels are measured when BH mass is roughly a) 0.04; b) 0.06; c) 0.08; d) 0.10. We also plot the fitting curve to these data points. Thick line represents the $\ln(N)$ dependence $\dot{N} = a\cdot \ln{N}+b$, thin line represents $N$ dependence $\dot{N} = a\cdot N+b$ with $a$ and $b$ as fitting constants.}
  \label{fig_Ndot_N_dependence_1e-3}
\end{figure}

To demonstrate the scaling relation for empty loss-cone, we make a plot showing the TDR dependence on $N$ in Fig. \ref{fig_Ndot_N_dependence_1e-3}. We use the same simulation data as in Fig.\ref{fig_Mdot}. However, we require in each panel the BH mass be roughly the same so that they are in the same evolutionary stage. In panel a), $M_{\bullet}$ in all 4 models is small, one can see that the data points are deviated from the $\ln{N}$ fitting curve, especially for 16K model. In the other 3 panels, however, BH is massive enough to be treated as a ``static" object in the cluster center and we see the $N$ dependence of TDR do follow the $\ln{N}$ form of Eq.\ref{Ndot_empty_lc} in empty loss-cone regime.

\begin{figure}[htbp]
  \begin{center}
  \includegraphics[width=\columnwidth]{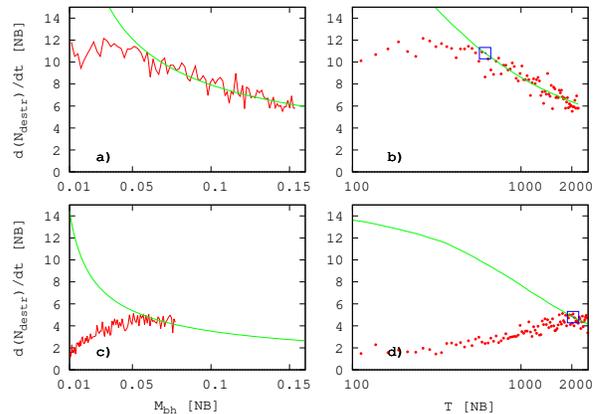}
  \end{center}
  \caption{TDR evolution with BH mass and integration time. Panel a) and c) show the TDR of 128K model as function of $M_{\bullet}$ together with the fitting curve. Panel b) and d) show the TDR as function of time, dots are measured data, curve is calculated value using the fitting parameters. Empty square marked the time when BH mass reaches $~0.06$. Panel a) and b) have $r_t=10^{-3}$; c) and d) have $r_t=10^{-4}$. }
  \label{fig_Mdot_mass_dependence}
\end{figure}

\cite{BME2004a} derived a formula to show how does TDR depend on BH mass, tidal radius, stellar mass and central density in empty loss-cone regime (Eq. 9 in their paper). Here we simply assume that $\dot{N} \propto r_t^{4/9} / M_{\bullet}^{\alpha}$ which is a mimic of their expression. And we also assume after a cusp is well established central density can be treated as a constant. Apply this relation to our data, we can fit the TDR as a function of $M_{\bullet}$ (Fig. \ref{fig_Mdot_mass_dependence}) with $\alpha = 0.669$ which is about $ 10\% $ higher than $11/18$ in their case. From panel a) and c) one can see that when BH mass is heavier than 0.06 ($\gamma = 1.27\cdot 10^{-4}$) the measured TDR agree with the analytical curve, which indicate at this point the system begins to enter the empty loss-cone regime. After getting the fitting parameters we calculate the analytical TDR using $M_{\bullet}$ at different evolution time, which are plotted in panel b) and d). In both panels one can see before BH mass reaches $\sim0.06$, TDR deviates largely from the theoretical curve which is based on the assumption of empty loss-cone and well established density cusp. In panel b) ($r_t = 10^{-3}$), BH grows faster and hence reaches 0.06 earlier than that in panel d) ($r_t = 10^{-4}$). After that point, simulation data points are well following the theoretical curve. \cite{BME2004a} uses King model with $W_0 = 10$ to construct their star cluster. Their model contains a small core and initial cusp outside the core. Central density in their runs is higher than ours during the whole simulation. In our case, cusp forms during the simulation and the final cusp is a little bit shallower than theirs. Despite the differences between their models and ours, the accretion behavior of BH in empty loss-cone regime agree with each other.

\begin{figure}[htbp]
  \begin{center}
  \includegraphics[width=0.8\columnwidth]{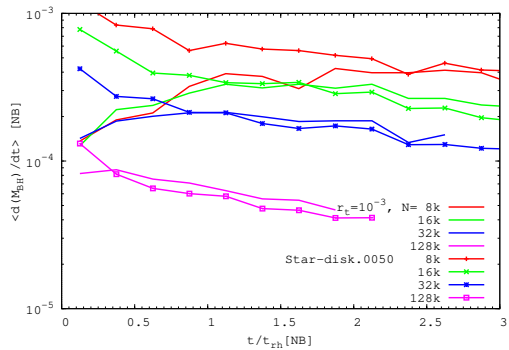}
  \end{center}
  \caption{Comparison between our work and \cite{JYM2012}. $x$ axis is time expressed in unit of initial half-mass relaxation time ($t_{rh}$), $y$ axis is the averaged disruption rate in a given time interval (i.e. 1/4 $t_{rh}$). The unit for disruption rate is mass increase per unit time.  Curves with symbols are results of \cite{JYM2012} , from top to bottom, correspond to $N$ = 8, 16, 32, and 128K. In their model $r_t = 1.1 \times 10^{-3}$. Curves without symbols shows our results for $r_t = 10^{-3}$, from top to bottom, correspond to $N$ = 8, 16, 32, and 128K.}
  \label{fig_Mdot_comp}
\end{figure}

To support our view how full and empty loss-cone are related to the SMBH's motion we also compare our data with those of \cite{JYM2012} in Figure~\ref{fig_Mdot_comp}. Their SMBH mass in cluster mass units is ten times larger than ours (0.1), so at the same particle number their $\gamma$ value is ten times smaller, and in addition to that they artificially fix the SMBH in the center. Except for the difference at the beginning, for models of same $N$ and $r_t$ our results agree with theirs in long term evolution. Since they have a fixed BH, loss-cone will be empty from beginning which is shown by the very good scaling of ${\dot M} \propto 1/N$ in the plot of their data. Our models start with smaller TD rate due to smaller initial SMBH mass, but after some time our TDR will catch up with theirs; also one can see that this ``catching up'' happens earlier for models with larger particle number. So we conclude that there is some self-regulation to reach a standard model of TDR, which only depends on the total particle number and tidal radius, but not on the initial mass of the SMBH.

\begin{figure}[htbp]
  \begin{center}
  \includegraphics[width=0.8\columnwidth]{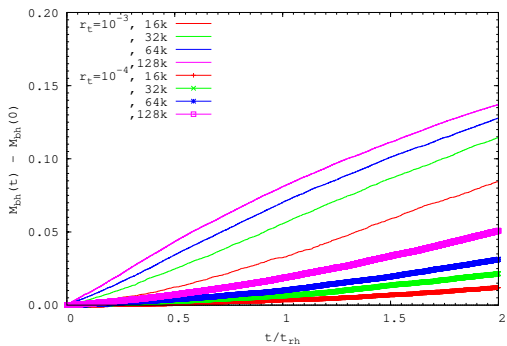}
  \end{center}
  \caption{Accreted mass as a function of time.}
  \label{fig_Mass-bh}
\end{figure}

The final SMBH mass is roughly 10 times the initial mass for the large value of $r_t = 10^{-3}$, and even in the case of the smaller tidal radius models, this factor is 2 - 5( see Fig~\ref{fig_Mass-bh}). These results should be interpreted with care if applied to real black holes and galactic nuclei. First, a scaling should be applied to realistic values of $r_t$ (which are usually much smaller than in our models) and of $N$ (in real galactic nuclei much larger than in our models); see for this analysis Sect.~\ref{Scaling}. Another issue is our assumption (which is quite common in the field, see e.g. also \citet{FR1976,BBK2011}) that a star is immediately disrupted as it crosses the tidal radius, and all of its mass is accreted to the SMBH. In reality some fraction of the stellar mass may escape from the SMBH, and the accretion of matter to the SMBH will be delayed for some orbital times, depending on the detailed disrupting process. The ejected mass is in the form of gas, which is not supported in our simulation code. Recent models of \citet{G_RR2012} may provide improved data on the mass fraction of tidal debris accreted to the SMBH and the ejected mass fraction, which we could use for an improvement of our models in the future.

\subsection{Density profile of the model cluster}

\begin{figure}[htbp]
  \begin{center}
  \includegraphics[width=\columnwidth]{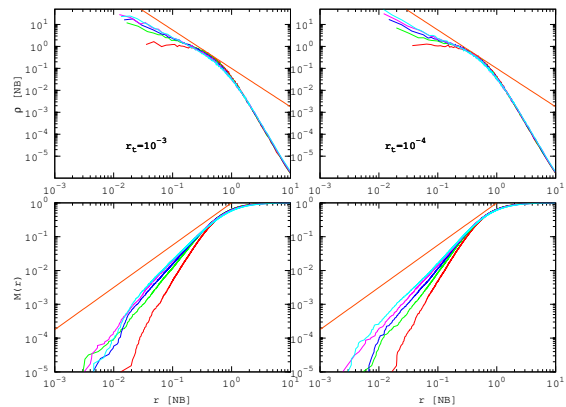}
  \end{center}
  \caption{Density profile evolution of model clusters. Plots in left column have $r_t=10^{-3}$, those in right column have $r_t=10^{-4}$. $x$ axis is the distance from the density center in N-body length units. $y$ axis is density in upper panels and normalized cumulative mass in lower panels. All the axes are in logarithmic scale. In each panel, curves from bottom to top indicate the density profile at $0.0, 0.5, 1.0, 1.5, 2.0 t_{rh}$. The cusp follows power law profile $\rho \propto r^{-\alpha}$. In the case of Bahcall-Wolf cusp $\alpha = -1.75$ which is also plotted here by straight line.}
  \label{fig_Mdot_1}
\end{figure}

In Fig.~\ref{fig_Mdot_1} we show the stellar density profile for our different models and how it evolves with time.

\begin{figure}[htbp]
  \begin{center}
  \includegraphics[width=0.8\columnwidth]{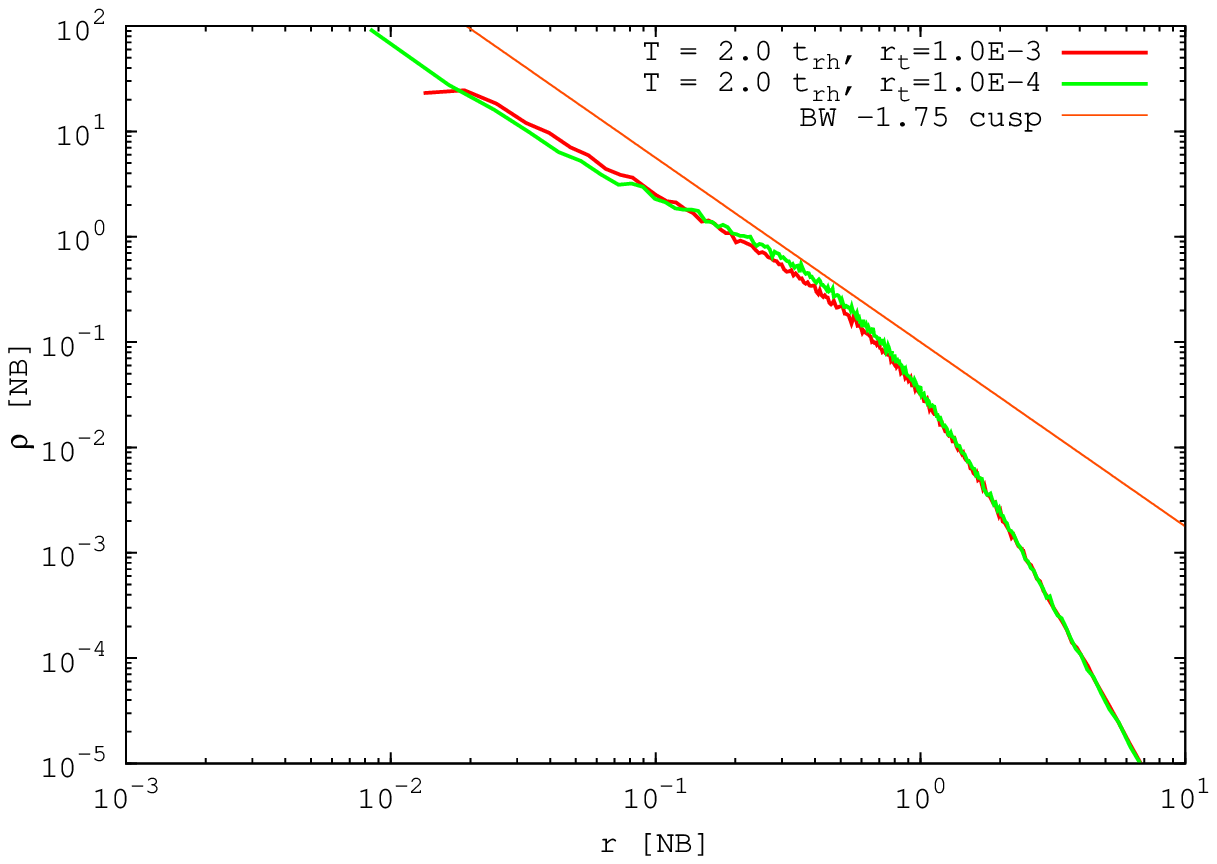}
  \includegraphics[width=0.8\columnwidth]{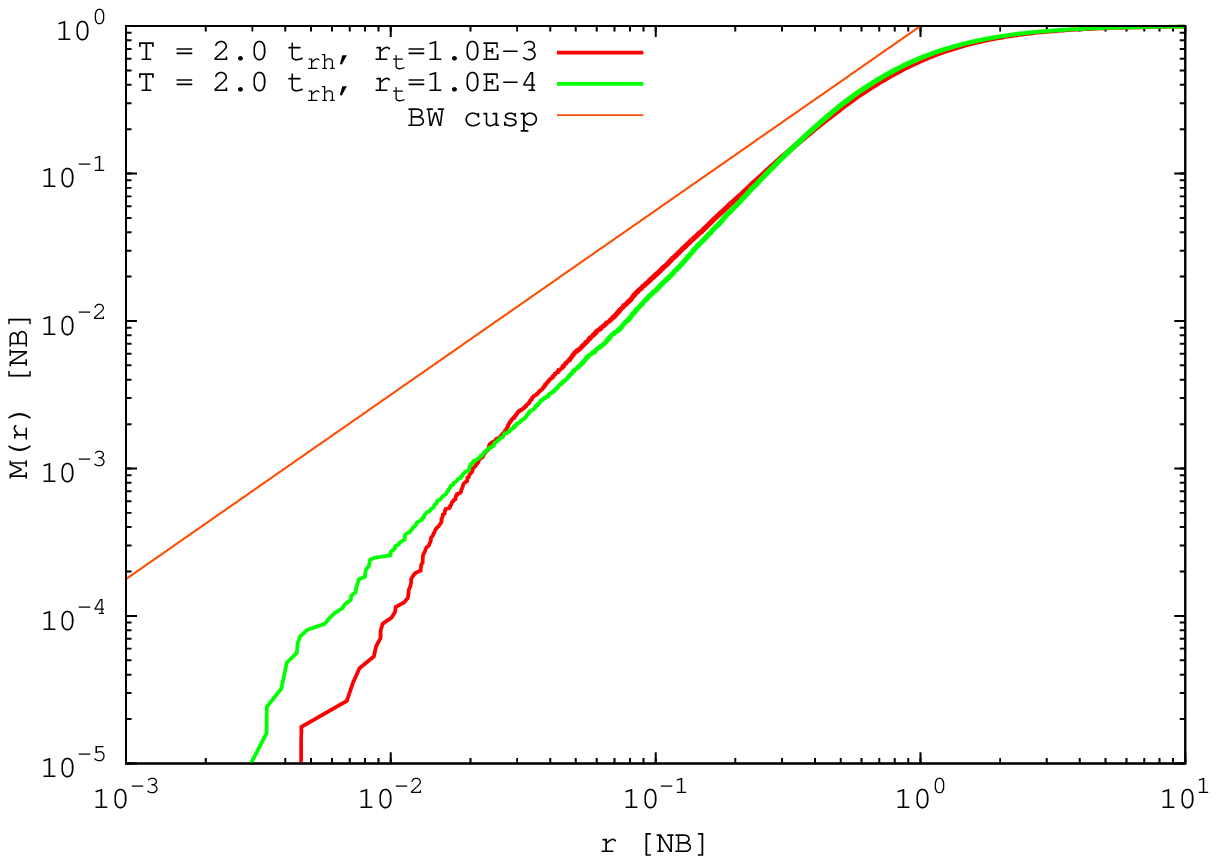}
  \end{center}
  \caption{Density profile for model N30 and N31. $x$ axis is distance from the density center in N-body length unit. $y$ axis is density in upper panel and normalized cumulative mass in lower panel. All the axis are in logarithmic scale. The cusp follows power law distribution $\rho \propto r^{-\alpha}$. In the case of Bahcall-Wolf cusp $\alpha = -1.75$ which is plotted as the red line.}
  \label{fig_Mdot_3x}
\end{figure}

Our simulations are starting with a Plummer model, which has a flat core in the initial density profile, and is not adjusted to the SMBH. Due to the presence of the SMBH, a density cusp quickly forms during the first 0.25$t_{rh}$. Afterwards it evolves slowly towards a steady state solution. We measured the slope of these cusps, none of them does exactly agree with the slope of Bahcall-Wolf cusp solution, which is $-1.75$. The final value in most of our simulation is around $-1.5$. An extreme case is $-1.69$ in model N31 achieved at 2 $t_{rh}$ (Figure~\ref{fig_Mdot_3x}). Although the SMBH's deep potential well pulls stars in, its vigorous motion, however, stirs up the central region of the cluster, playing an opposite role in the cusp formation process.

%%%%%%%%%%%%%%%%%%%%%%%%%%%%%%%%%%%%%%%%%%%%%%%%%%%%%%%%%%%%
\subsection{Origin of disrupted stars}

One important concept in loss-cone theory is the critical radius ($r_{crit}$), where the star removal process is just balanced with its repopulation. It has been first defined by \citet{FR1976}, and is typically large compared to the Schwarzschild radius or the stellar tidal disruption radius of a SMBH. \citet{FR1976} show that in many cases the critical radius is even larger than the gravitational influence radius $r_h$ of the SMBH (implying that most of the stars getting accreted are unbound), but for SMBH masses larger than a certain value $r_{crit}$ can become smaller than $r_h$. Only in those latter cases can we expect a Bahcall-Wolf density cusp to form. Here we want to find out empirically whether the definition of $r_{crit}$ is useful and whether it can be reproduced using our simulation data.

We use the following procedure to find $r_{crit}$ empirically: first, we determine the apocenter of every disrupted star from its energy, angular momentum, and the gravitational potential in the star cluster. Here we neglect that a small fraction of the stars may have been scattered into the loss-cone by two-body relaxation inside $r_{crit}$. Then we look at the distribution of apocenter distances (denoted as $R_{max}$) of all tidally accreted stars, by a radial binning procedure.

\begin{figure}[htbp]
  \begin{center}
  \includegraphics[width=0.8\columnwidth]{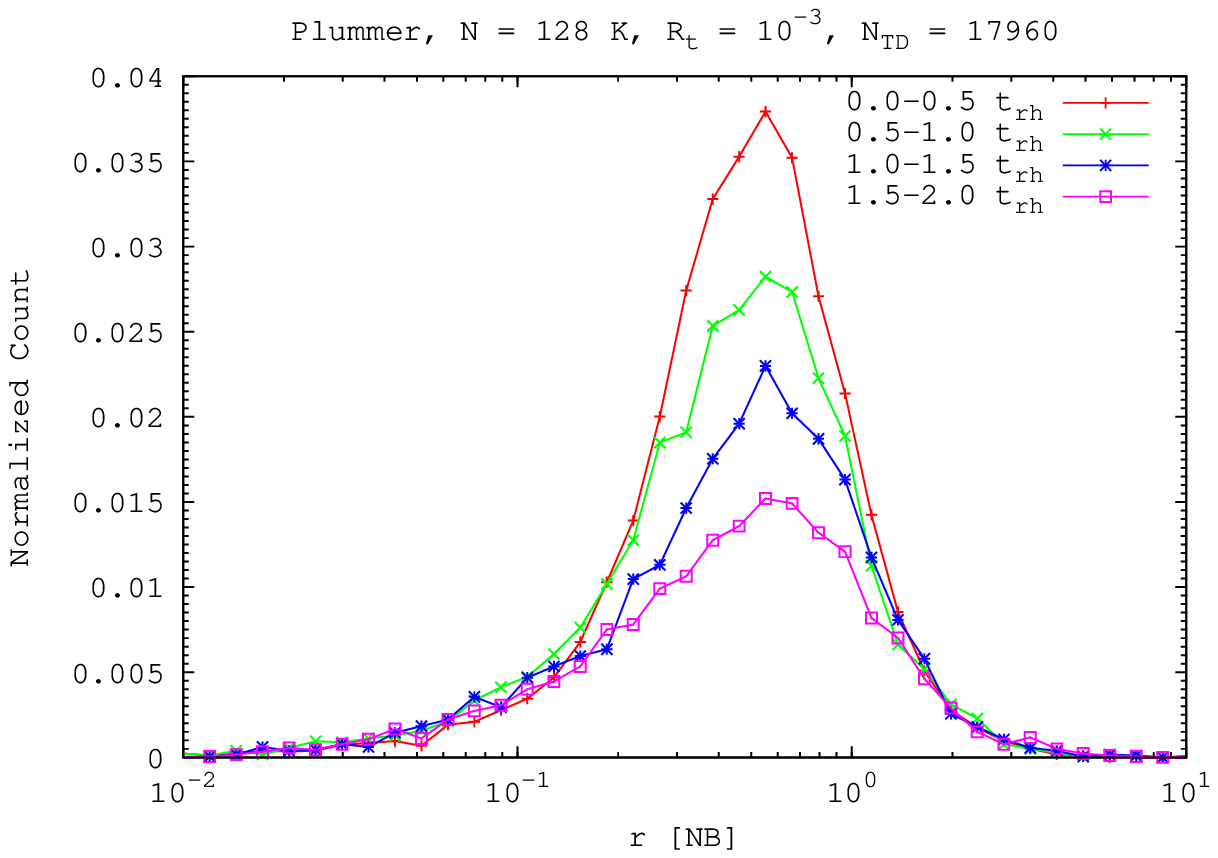}
  \includegraphics[width=0.8\columnwidth]{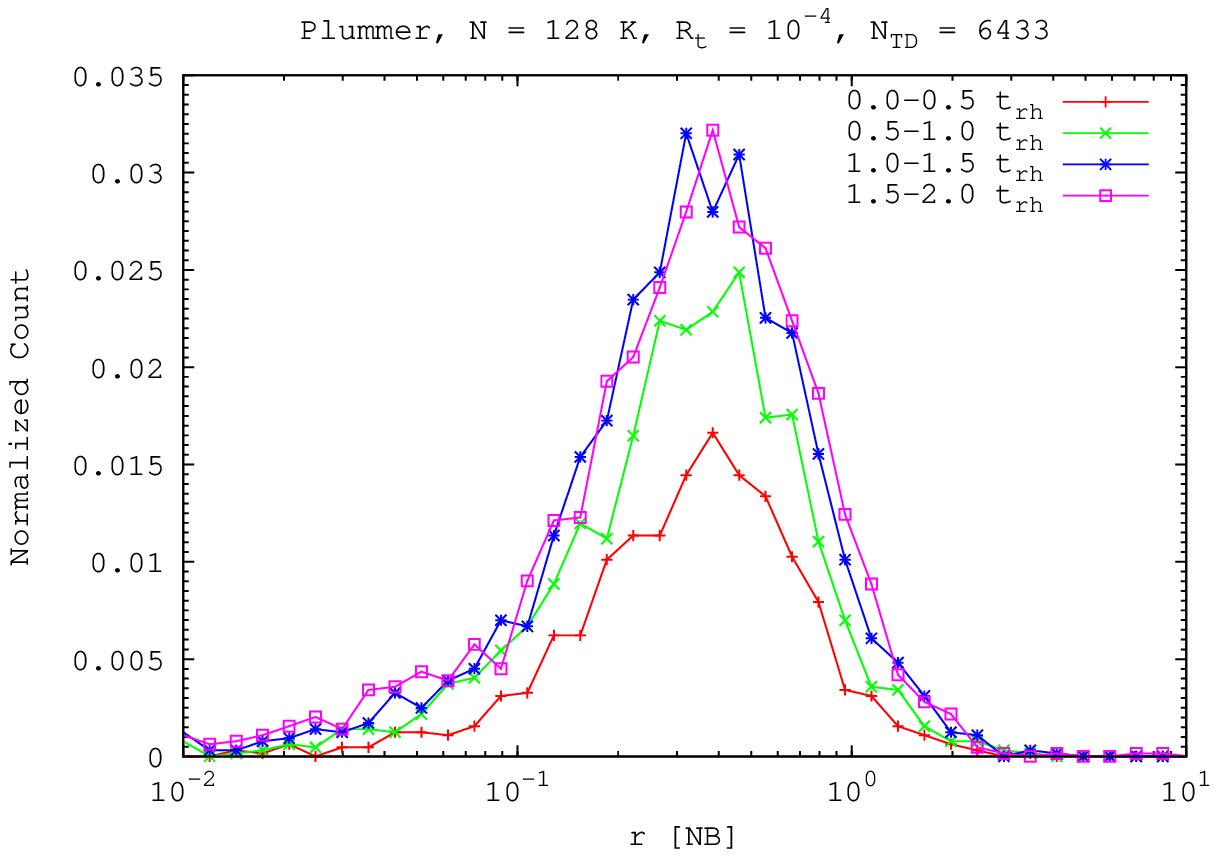}
  \end{center}
  \caption{$R_{max}$ distribution. $x$ axis is distance from the density center in N-body length unit. $y$ axis is normalized count in bins, normalized by the total number of disrupted stars through out the run. Both plots are from models containing 128K particles. Top: plot for $r_{t}=10^{-3}$. Bottom: plot for $r_{t}=10^{-4}$.}
  \label{fig_Rmax31}
\end{figure}

Fig~\ref{fig_Rmax31} shows the $R_{max}$ distribution for models of $r_{t}=10^{-3}$ (N30) and $r_{t}=10^{-4}$ (N31). There is a clear peak of the distribution, its location is nearly independent of time and is far from the BH. Also the peak position for $r_{t}=10^{-3}$ is larger than that for $r_{t}=10^{-4}$. These peaks resemble that expected in loss-cone theory, but before making any conclusion we need to check it more carefully.

\begin{figure}[htbp]
  \begin{center}
  \includegraphics[width=0.8\columnwidth]{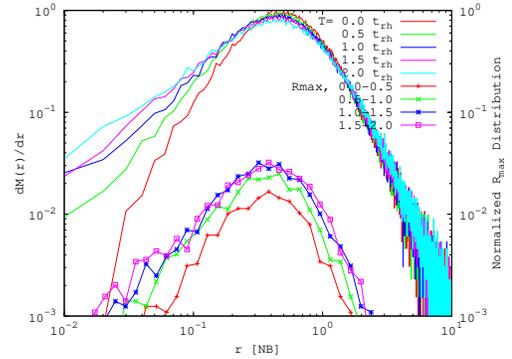}
  \end{center}
  \caption{$R_{max}$ and shell mass distribution. $x$ axis is distance from the density center in N-body length unit. Left $y$ axis is shell mass $dM(r)/dr$, where $M(r)$ is the enclosed mass at $r$. Right $y$ axis is normalized count for $R_{max}$ distribution. The upper curves are shell mass profile, lower curves are $R_{max}$ distribution. Please note in this plots the $y$ axis is logarithmic scale and normalization procedure is different from former plots, so $R_{max}$ curves looks different.}
  \label{fig_Mshell-Rmax}
\end{figure}

Fig~\ref{fig_Mshell-Rmax} shows the shell mass ($dM(r)/dr = 4 \pi r^2 \rho$) distribution together with the normalized $R_{max}$ distribution for comparison. We will use these plots to explain the origin of the peaks in $R_{max}$ distribution. There are only a few TD stars coming from small radii, as is expected from theory in case of an empty loss-cone. However, the shell mass is also small at central region. So the small value in $R_{max}$ distribution should be a consequence of the low star numbers in this region. Both $R_{max}$ and shell mass curves have a peak at some specific radii. Are the positions of these two peaks the same? If they are, then the drop beyond the peak in $R_{max}$ distribution might caused by the decrease in star numbers in shells and is difficult to distinguish from the prediction of loss-cone theory.

\begin{figure}[htbp]
  \begin{center}
  \includegraphics[width=0.8\columnwidth]{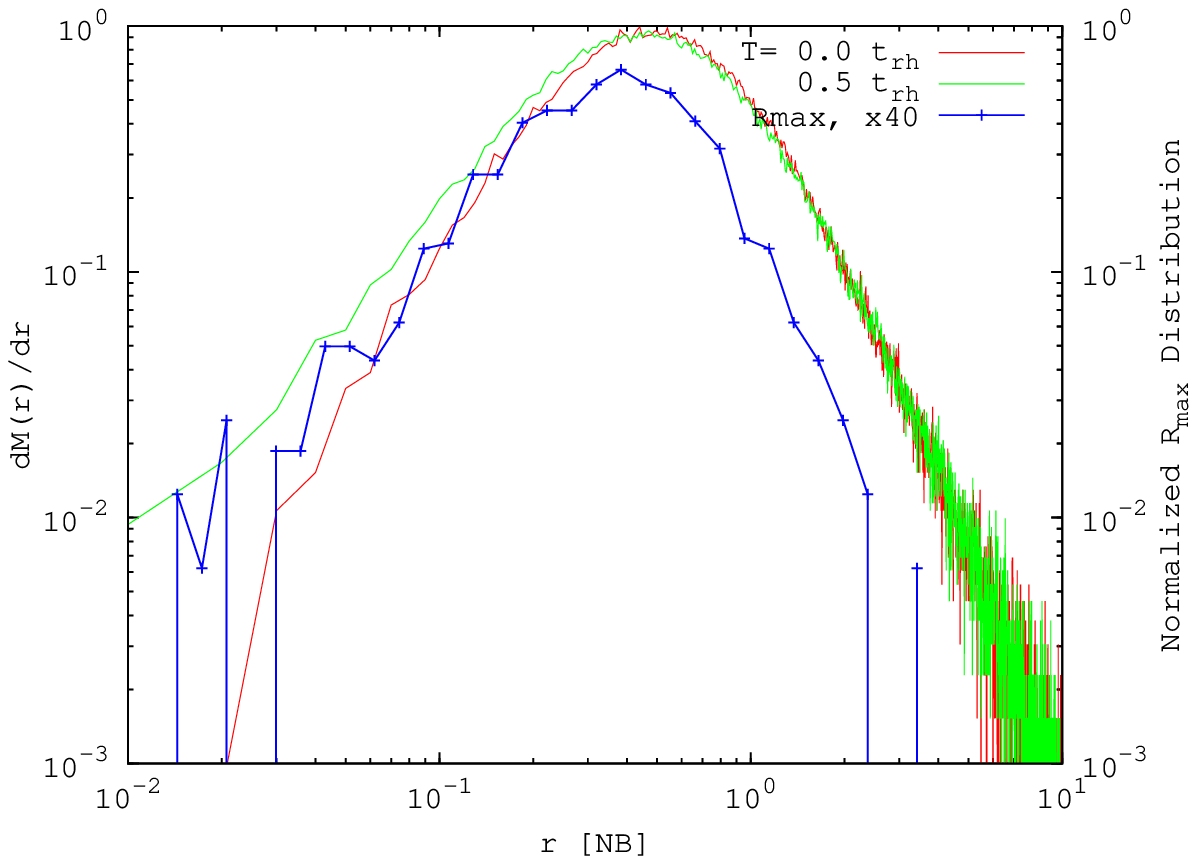}
  \includegraphics[width=0.8\columnwidth]{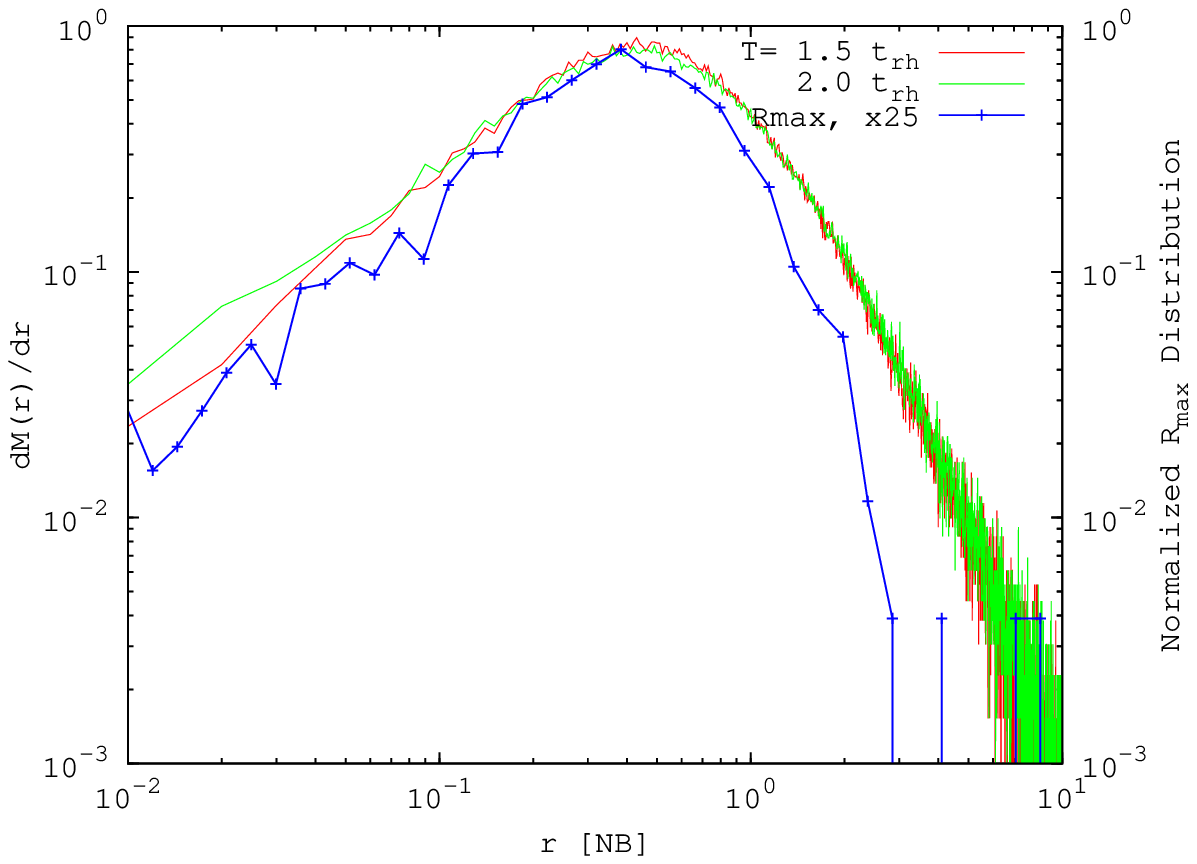}
  \end{center}
  \caption{Shell mass profile together with normalized $R_{max}$ distribution for $r_t=10^{-4}$ model. Top panel is measured in the first $0.5 t_{rh}$ and bottom is in the last $0.5 t_{rh}$. $x$ axis is distance from the density center in N-body length unit. $y$ axis is shell mass. From top to bottom are different time intervals. For the purpose of comparison, data points of $R_{max}$ are multiplied by 40 (top) and 25 (bottom).}
  \label{fig_Mshell-Rmax31-x}
\end{figure}

\begin{figure}[htbp]
  \begin{center}
  \includegraphics[width=0.8\columnwidth]{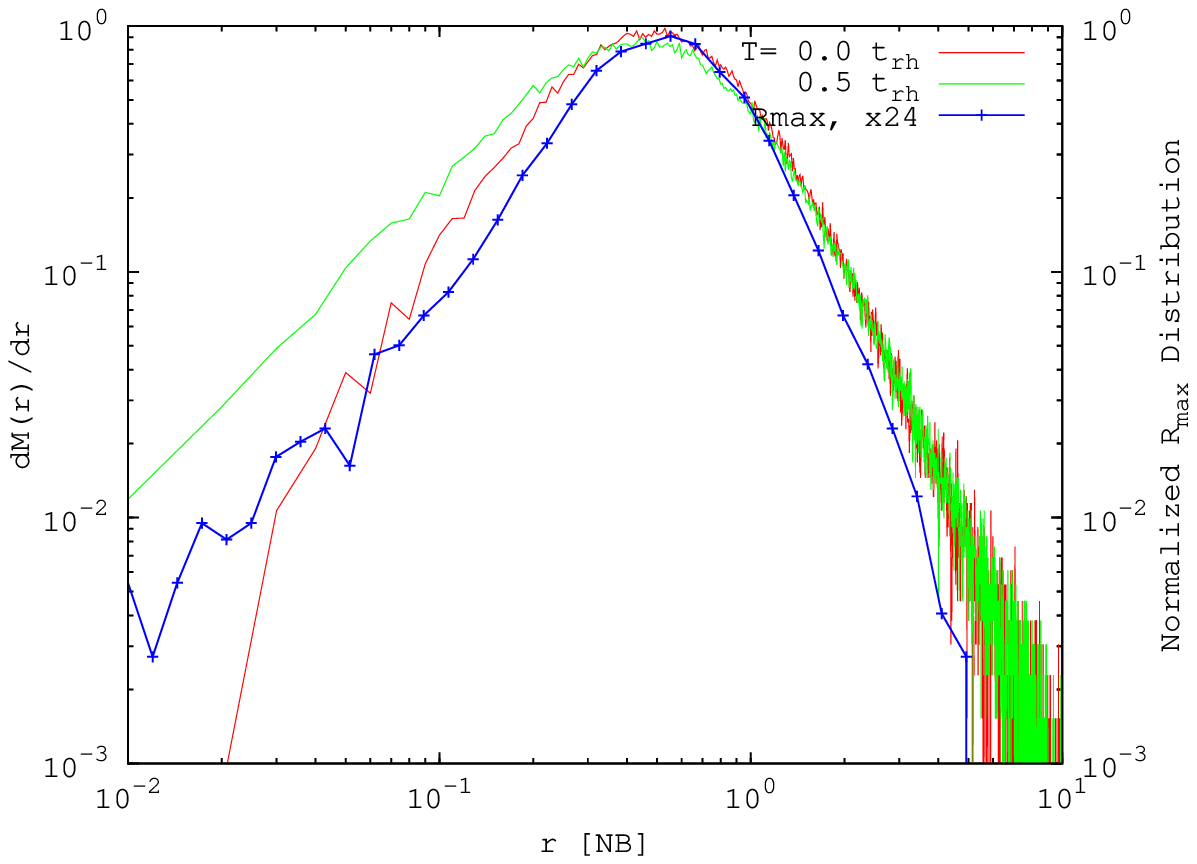}
  \includegraphics[width=0.8\columnwidth]{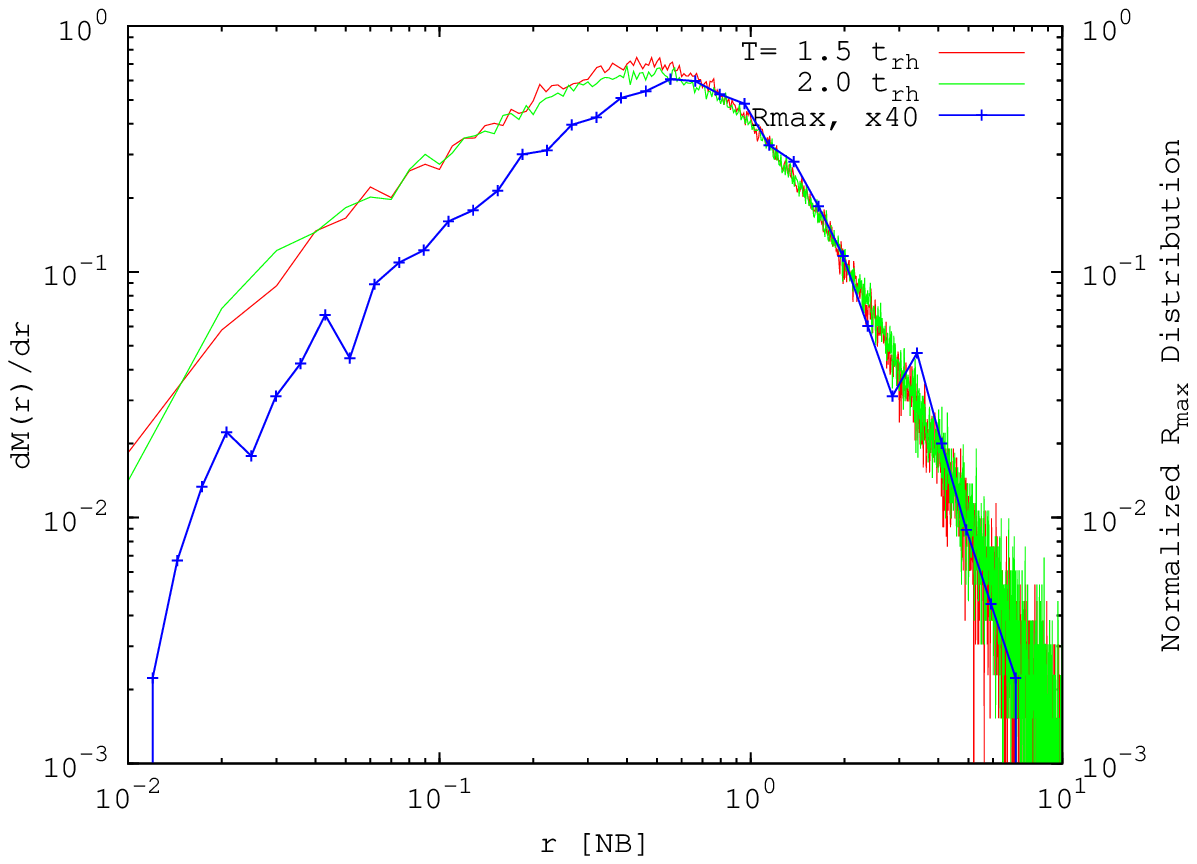}
  \end{center}
  \caption{These plots show shell mass profile together with normalized $R_{max}$ distribution for $r_t=10^{-3}$ model. Top panel is measured in the first $0.5 t_{rh}$ and bottom is in the last $0.5 t_{rh}$. $x$ axis is distance from the density center in N-body length unit. $y$ axis is shell mass. From top to bottom are different time intervals. For the purpose of comparison, data points of $R_{max}$ are multiplied by 24 (top) and 40 (bottom).}
  \label{fig_Mshell-Rmax30-x}
\end{figure}

In order to see whether this is the case, we move the $R_{max}$ curves upward and let the left part overlap with the shell mass curve (see Fig~\ref{fig_Mshell-Rmax31-x}). One can see the slope of $R_{max}$ curves follow $dM/dr$ profile at small radii. The maximum number of stars in the loss cone to be disrupted can be estimated from its size $\Omega$, defined in every mass shell by the surface area of a cone with opening angle $\theta_{lc}$ relative to the full solid angle of a sphere, which represents all stars at this radius; it is $\Omega = \theta_{lc}^2 / 4 $ (note we have used here the notation of \cite{AFS2004}).

For the models with small tidal radius ($10^{-4}$) we see in Fig.~\ref{fig_Mshell-Rmax31-x} that the maximum $R_{max}$ for the tidally accreted stars is reached at significantly smaller radii than the location of maximum of the mass distribution of all stars. In the upper panel while the shell mass keeps on increasing, $R_{max}$ curve begins to fall at a specific radius. However, for model with large tidal radius (Fig.~\ref{fig_Mshell-Rmax30-x}) the maximum and (for larger radii) subsequent drop of the $R_{max}$ curve coincides with that of the shell mass curve. While in the first case we can conclude there is a critical radius at the peak position of $R_{max}$ distribution, the peak of $R_{max}$ in the second case only reflects the steep decrease of density (maximum of shell mass profile), and does not give clear evidence, where is the location of $r_{crit}$, defined as the place where $\theta_{lc} = \theta_D$. Outside of $r_{crit}$ loss-cone loses its significance, stars can get into and out of it freely so that contribution from outside regions vanishes.

%%%%%%%%%%%%%%%%%%%%%%%%%%%%%%%

\subsection{Orbital eccentricity of the disrupted stars}

When a TD event happens, after the peak luminosity the light decline can last for months to years, depending on the way it is disrupted. So the distribution of orbital parameters of the disrupted stars is important to know for the observational counterparts of TD ~\citep{GRR2013}.

\begin{figure}[htbp]
  \begin{center}
  \includegraphics[width=0.8\columnwidth]{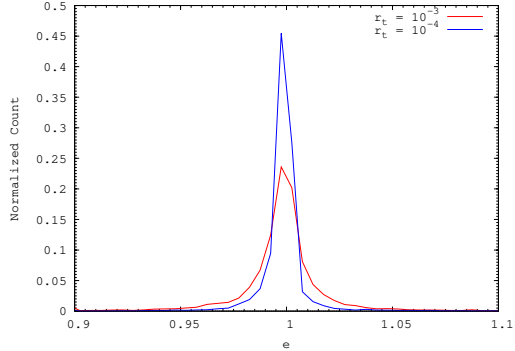}
  \end{center}
  \caption{Eccentricity distribution. $x$ axis is eccentricity. $y$ axis is normalized count.}
  \label{fig_ecc}
\end{figure}

Loss-cone theory predicts that stars, which have larger energy than $E_t$ should have very small angular momentum so that they can get close enough to black hole. These stars are expected to move on very eccentric orbits. In order to check this directly and quantitatively, we measured the TD stars' orbital eccentricity using the Runge-Lenz vector, with the central black hole as a reference point:

\begin{equation}
\mathbf{e} = \frac{ \mathbf{v} \times \mathbf{J} }{ G(M_{\bullet}+m)} - \frac{\mathbf{r}}{r} \,
\label{Eq_Runge-Lenz}
\end{equation}

\noindent where $\mathbf{v}$ is the relative velocity between the BH and the star, $\mathbf{r}$ is their relative distance, and $\mathbf{J}=\mathbf{r} \times \mathbf{v}$ is the relative angular momentum per unit mass. Two examples of our results are plotted in Fig~\ref{fig_ecc}. Almost all the stars are concentrated around $e=1$ (parabolic orbit marginally bound), roughly half of them have $e$ larger than 1, indicating they are unbound to the black hole. Because the black hole is moving, the Runge-Lenz vector even for bound stars is not strictly constant. So as the integration goes on we get some bound stars unbound and vice versa. Here we just take the last measurable value before TD events. Stars changes their energy also by interactions with other stars, but we do not directly measure this effect. What is the relation between $e$ and $R_{max}$ for every single accreted star? It is interesting to know the origin of tidally accreted stars in the surrounding galactic center.

\begin{figure}[htbp]
  \begin{center}
  \includegraphics[width=0.8\columnwidth]{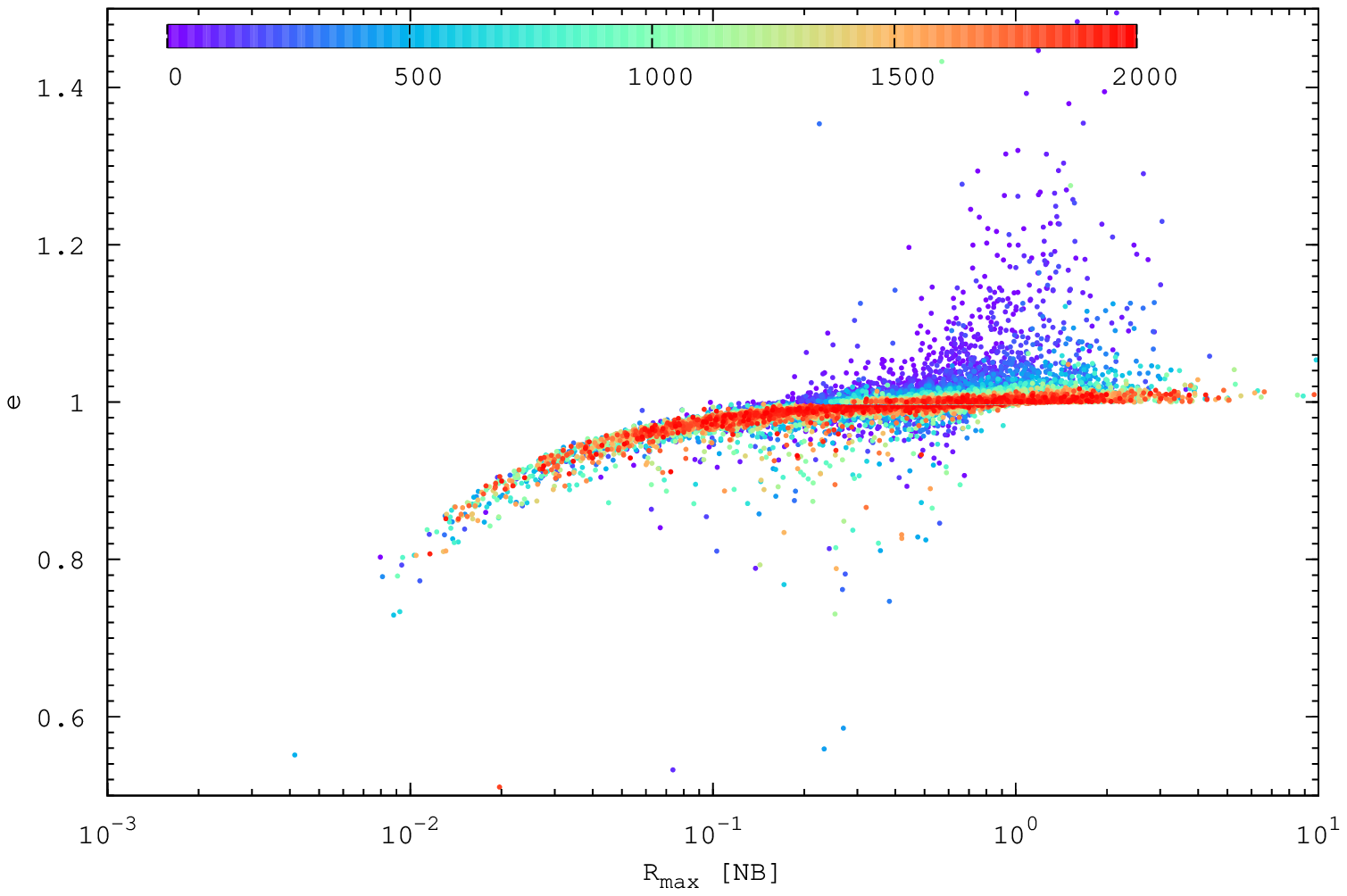}
  \includegraphics[width=0.8\columnwidth]{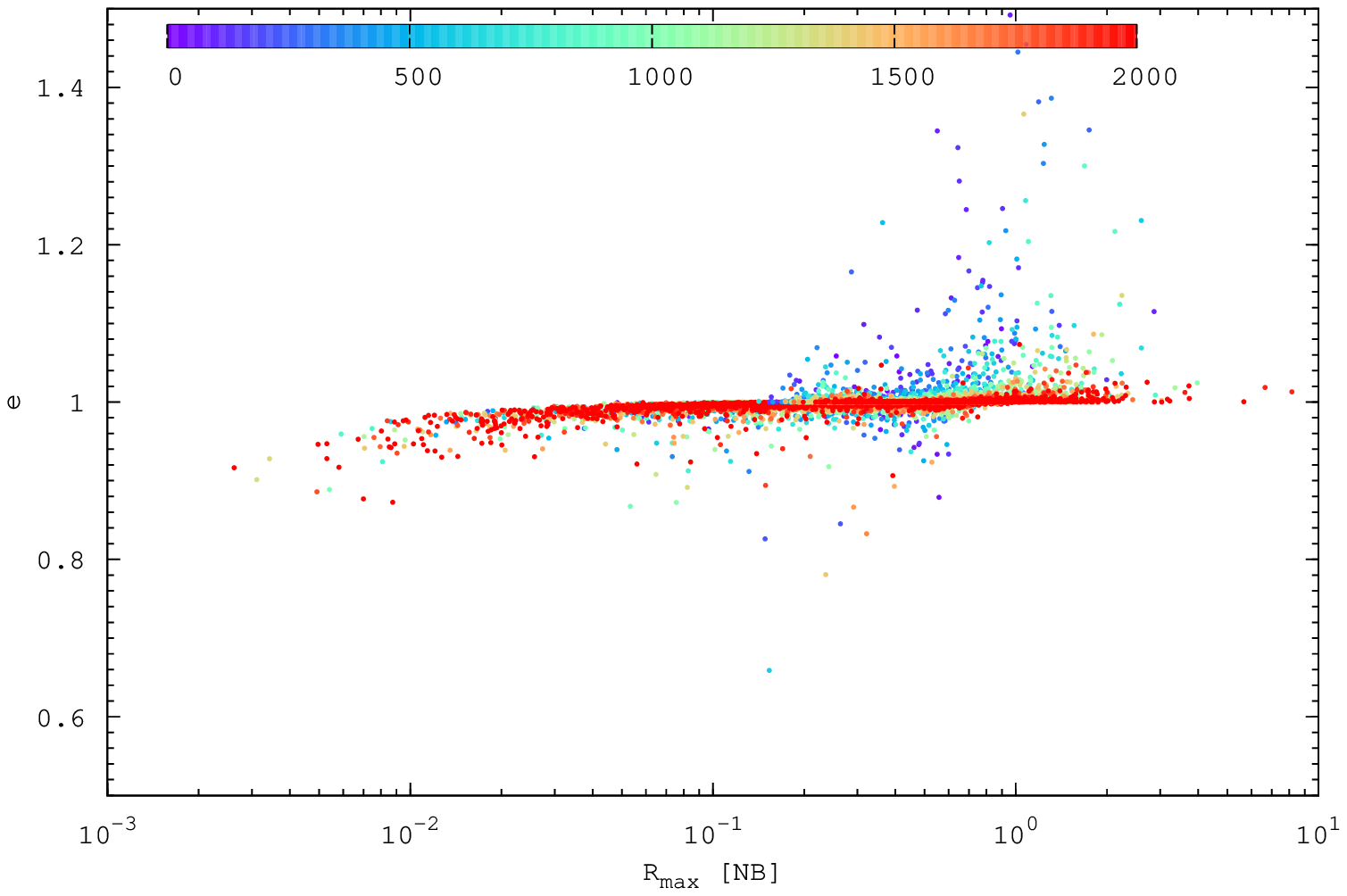}
  \end{center}
  \caption{Relation between eccentricity and $R_{max}$. $x$ axis is distance to BH. $y$ axis is eccentricity. Color bar indicate the time when star is disrupted. Top: $r_t=10^{-3}$. Bottom: $r_t=10^{-4}$}
  \label{fig_eccXRmax_with_time}
\end{figure}

Fig~\ref{fig_eccXRmax_with_time} shows the relation between $e$ and $R_{max}$ for individual stars at the time of disruption. In the plots we see a special position inside which all stars have $e\leqslant1$. For $r_t=10^{-4}$ model the distance of this critical point to the center is smaller than that in $r_t=10^{-3}$ model. We also note that within this critical point, the 2 models behaves differently. An upper boundary of $e$ is clear to see in both plots, and the shape of this boundary varies with $r$. In $r_t=10^{-3}$ model this boundary drops quickly toward center. It already drops to 0.8 at $r=0.01$ while in $r_t=10^{-4}$ model this value is still close to 1. Furthermore, we add time information to these plots in Fig.~\ref{fig_eccXRmax_with_time} and find these phenomenon are connected to the black hole's Brownian motion (see Fig.~\ref{fig_Ndot_bhxv}); in early stage the amplitude of Brownian motion is large, the accreted unbound stars have a large dispersion of eccentricities (points with color from purple to light blue in Fig.~\ref{fig_eccXRmax_with_time}); once it has decayed below the threshold, i.e. the black hole comes to ``rest", we are close to the classical loss-cone case where all stars are accreted with eccentricities very close to unity (points with color from green to red).

Here we see the choice of $10^{-3}$ as tidal radius is really large. BH with this $r_{t}$ can destroy a large number of stars initially, which intersect BH's path by chance due to its large initial displacement from the center. For small $r_t$ models the situation is better and behaves more like theoretical expectation of classical loss-cone.

%%%%%%%%%%%%%%%%%%%%%%%%%%%%%%%

%%%%%%%%%%%%%%%%%%%%%%%%%%%%%%%

\subsection{Velocity dispersion}

As we have seen in the former subsection, most of the disrupted stars are moving on nearly radial orbits (provided the tidal radius is not too large and the black hole motion has sufficiently decayed to establish the classical loss cone case). When these stars are disrupted on their radial orbits, the velocity dispersion of the surrounding star cluster will become tangential biased. We measured the anisotropy parameter $\beta$ throughout the whole model cluster to see how it changes with time as a consequence of tidal disruption. $\beta$ is defined as

\begin{equation}
\beta = 1 - \frac{\sigma_{t}^{2}}{2\sigma_{r}^{2}}
\label{Eq_beta}
\end{equation}

\begin{figure}[htbp]
  \begin{center}
  \includegraphics[width=0.8\columnwidth]{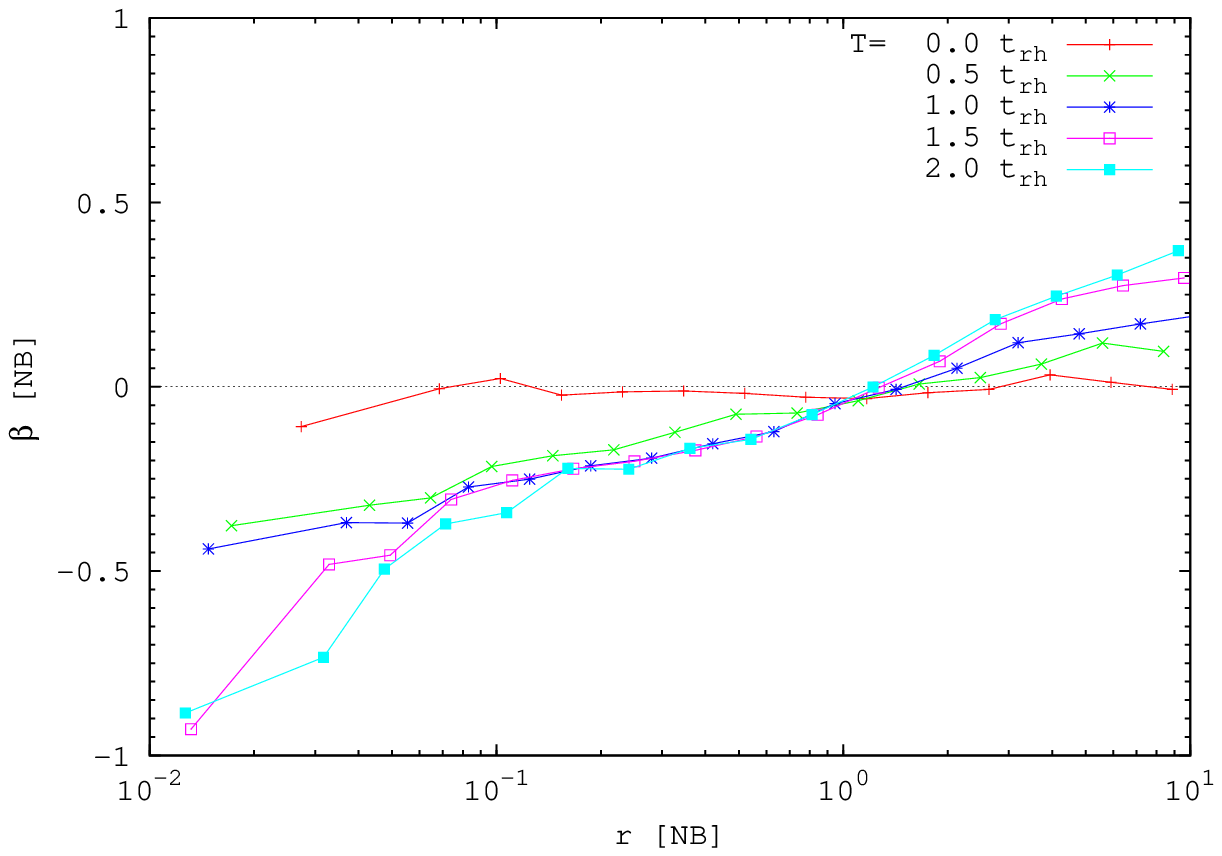}
  \includegraphics[width=0.8\columnwidth]{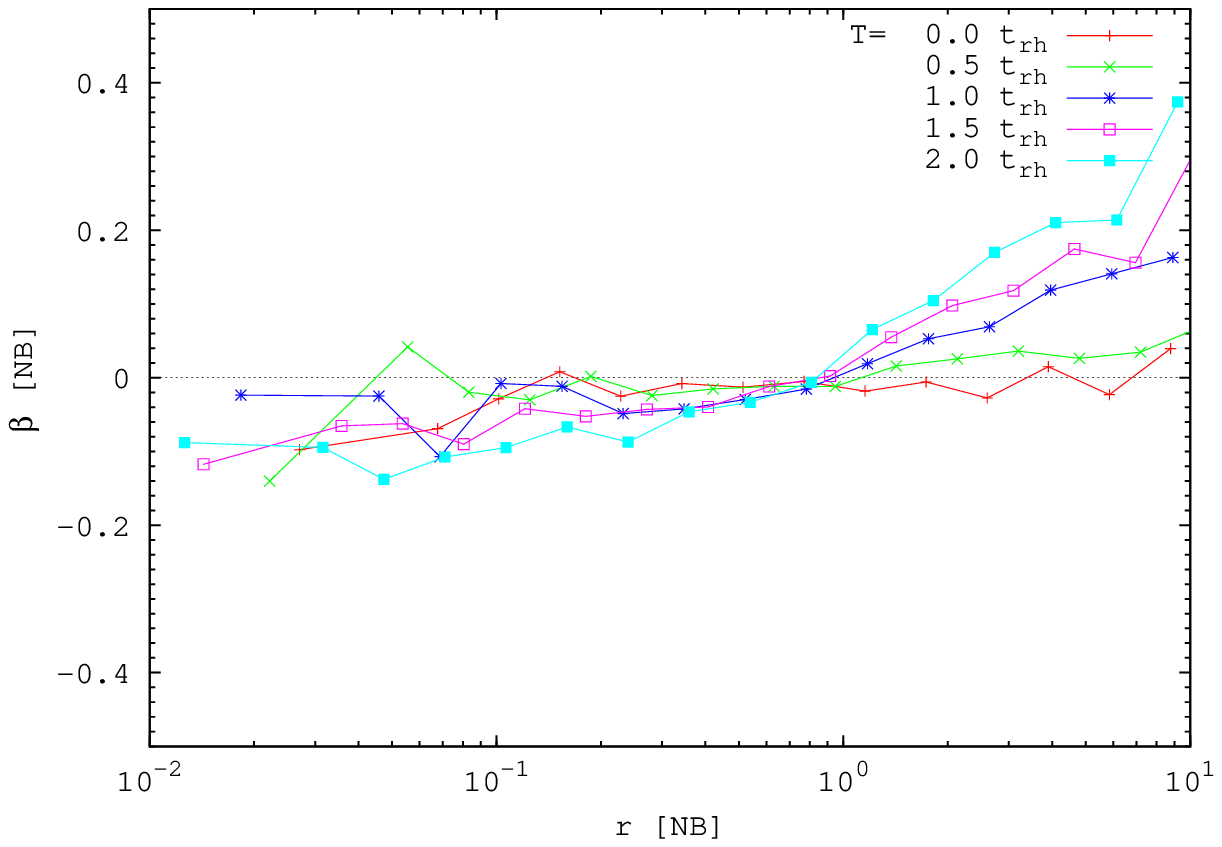}
  \end{center}
  \caption{Anisotropic parameter $\beta$ as a function of $r$. $x$ axis is distance from BH. $y$ axis is $\beta$. Top: $r_t=10^{-3}$. Bottom: $r_t=10^{-4}$}
  \label{fig_beta}
\end{figure}

\noindent According to this definition, $\beta < 0$ means velocity dispersion is tangential; $\beta > 0$ means dispersion is radial; otherwise we have an isotropic distribution (note that some stellar dynamical papers prefer to define the quantity $A = 2\beta$). $\beta$ can vary between unity (only radial orbits) and $-\infty$ (only circular orbits). Larger loss-cone will result in larger anisotropy if the loss-cone is empty. The value of $\beta$ can somehow indicate the status of loss-cone. Fig~\ref{fig_beta} shows $\beta$ curve for 2 set of $r_t$ and also their evolution. As one can see, large $r_t$ model shows a continuously decreasing $\beta$ in the inner region of the cluster. For small $r_t$ model this trend is less pronounced, but still clearly visible. This could be explained by the empty loss-cone in velocity space. However, in the case of $r_{t} = 10^{-3}$, the large deviation from isotropy is more than an empty loss-cone could do. Thus the large amplitude of Brownian motion at early stage somehow produced an enlarged loss-cone, or we can refer it as effective loss-cone~\citep{LT1980}. Stars enters this effective loss-cone can be tidally destroyed by a probability $P$, but this $P$ is less than 1. So the moving black hole digs a larger hole in the velocity space than that can be done by a static black hole. $P$ may depend on $r_t$, namely larger $r_t$ results in larger $P$. So in the $r_t=10^{-4}$ model, the enlarged effective loss-cone is still full and we see $\beta$ only changes a little. In the outer region, we see an increasing $\beta$, this can be produced by purely a two-body relaxation effect ~\citep{BS1986,AFS2004}.

%%%%%%%%%%%%%%%%%%%%%%%%%%%%%%%

\subsection{Scaling to the Galactic Center}
\label{Scaling}

Our simulation results allow a prediction of tidal disruption (TD) rates in real galactic nuclei, especially the one of our own Milky Way. Since we do not directly model the entire galaxy, only consider the supermassive black hole (SMBH) and its surrounding dense stellar system (stellar mass up to hundred times the SMBH mass), a few steps are necessary to apply our results for real systems. Model $N$-body units used throughout this paper need to be transformed into physical units, a scaling from the usually smaller number of particles in the simulation to a realistic particle number $N$ is necessary. We first consider the scaling of the units. The observed $\rm M_{\bullet}$-$\sigma_{*}$ relation between the SMBH mass and the velocity dispersion of its host bulge \citep{FM2000,SG2011} is used to link the radial scale in a galaxy to our model units:

\begin{equation}
\lg(M_{\bullet}/M_{\odot}) = 8.18 + 4.32 \lg(\sigma_{*}/200 {\rm ~km/s})
\label{Eq_M_sig}
\end{equation}

\noindent With the definition of the influence radius $r_h$

\begin{equation}
r_h = \frac{GM_{\bullet}}{\sigma^{2}}
\label{Eq_influence_radius}
\end{equation}

\noindent one can get an expression connecting the SMBH mass and its influence radius in physical units for a real galaxy \citep{BBK2011}:

\begin{equation}
r_h = 1.09(\frac{M_{\bullet}}{10^{6}M_{\odot}})^{0.54} {\rm ~pc}
\label{Eq_Rinf_Mbh}
\end{equation}

Assume a BH of mass $10^{6} M_{\odot}$ inhabits the center of a target galaxy, we get $r_h = 1.09$ pc; in our model we use the definition $M(r_h) = M_{\bullet}$ and find $r_h = 0.1 [L]$, where $[L]$ is the $N$-body length unit. So we have $[L] \approx$ 10.9 pc. Similarly we compute the $N$-body time unit $[T] \approx 5.39 \times 10^4$ yr using $T^2 = R^3/(GM)$, at $r_h$.

In our simulations the maximum particle number is of the order of $10^5$, while in the real galactic environment the number of stars is many orders of magnitude higher, which is why we need an extrapolation. We have a cluster mass in our model of $10^8 ~M_{\odot}$, since the SMBH's mass is 1 percent of it. If further assume the average mass of the stars to be 1 solar mass, then the cluster contains $10^8$ stars. In Section~\ref{TDR} we see 2 types of $N$ dependence responsible for full and empty loss-cone regime, explicitly. Here we adopt the results in the empty loss-cone regime, which occurs in the later stages of our simulation, when we have a cusp and small Brownian motion of the SMBH. This compares well with the situation in our Galactic Center, because the central SMBH moves very little \citep{RB2004}.

\begin{figure}[htbp]
  \begin{center}
  \includegraphics[width=0.8\columnwidth]{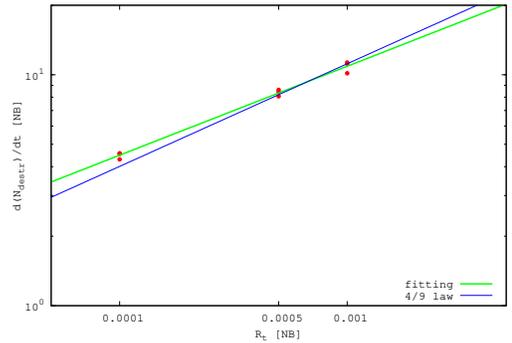}
  \end{center}
  \caption{TDR for different $r_t$ in empty loss-cone regime. Adopted data are taken from model N30, N31 and N32. For every $r_t$, the 3 data points are 3 largest values of TDR in each model and the corresponding SMBH mass is roughly the same. Thick line is power law fitting to the data with index 0.39. Thin line indicates the theoretical power law index 4/9.}
  \label{fig_Ndot_Rt}
\end{figure}

For the scaling to the Galactic Center we have now three steps: first use Eq.~\ref{Ndot_empty_lc} to scale the accretion rate as a function of particle number. We take one of our measured TDR, in model N30 ($r_t=10^{-3}$, $N=128K$), $\dot N$ equals 11.4 per unit time (see Fig ~\ref{fig_Mdot}, at $T = 0.5 t_{rh}$). This can be scaled up to to 19.3 per unit time when $N=10^8$. Second, we convert this in physical units as explained above, to get $3.58\times10^{-4}$ yr$^{-1}$. Third and finally, we have to scale to the real size of the tidal radius: $r_t$ for a $10^6 M_{\odot}$ SMBH is about $4\times10^{-7}$ $[L]$, which is roughly 4 orders of magnitude smaller than our adopted value. From the simulation we find at the stage we discussed here, $r_t$ dependence satisfy a power law relation $\dot{N} \propto r_t^{l}$, with $l = 0.39$ (See Fig.\ref{fig_Ndot_Rt}). Finally we get a TDR of $1.69 \times 10^{-5}$ yr$^{-1}$.

We also do the same procedure for larger SMBH, and find a decreasing TDR. This may mostly be caused by the converting of time unit (i.e. in $M_{\bullet} = 10^7 ~M_{\odot}$ case, $[T] = 1.1 \times 10^5$ yr; in $10^8 ~M_{\odot}$ case, $[T] = 2.2 \times 10^5$ yr), while $\dot N$ does not change much in the scaling. More details can be found in the appendix. Using our scaling formula (\ref{Eq_D06b}) to extrapolate our result into the $r_t$ range and particle numbers used by \cite{BME2004a}, we can recover their results in the sense of order of magnitude.

Applying the above procedure to our Milky Way center, where there is a $4.07 \times 10^{6} ~M_{\odot}$
SMBH \citep{G2009}, results in a TDR of $1.09 \times 10^{-5}$ yr$^{-1}$. In real galactic center there are many stellar species, such as main sequence star , RGB star, white dwarf (WD) star, neutron star (NS) and stellar-mass BH. In our current simulation, since all star particles have same mass and no stellar evolution process. It is hard to tell TDR for different stellar species. But we can make a simple estimate based on the fraction of each stellar species, if we ignore the effects introduced by stellar mass spectrum (e.g. mass segregation, different tidal radius, etc.). We assume a Kroupa IMF \citep{Kroupa2001} in mass range $0.08 - 100 M_{\odot}$, with solar metallicity and all stars are formed in one star formation epoch. At current evolution time ($\sim10 \rm{Gyr}$), we expect $10\%$ of total stars are WDs, $0.35\%$ are NSs and $0.16\%$ are stellar-mass BHs. Thus the accretion rate for these species would be $1.09 \times 10^{-6}$ yr$^{-1}$ (WD), $3.90 \times 10^{-8}$ yr$^{-1}$ (NS) and $1.74 \times 10^{-8}$ yr$^{-1}$ (stellar-mass BH). Mass spectrum and stellar evolution will be included in our future work.

Finally, we would like to point out the differences between our scaling formula from \cite{BME2004a}. Our scaling procedure to estimate TDR in real galaxies is based on the same physical principle than in theirs, i.e. Eq. 9 in \cite{BME2004a}. While they keep 4 parameters (stellar radius and mass, central density and $M_{\bullet}$) in their final formula, we reduce every thing to the dependence only on $M_{\bullet}$, by using some fiducial scaling relations (see Appendix).

%%%%%%%%%%%%%%%%%%%%%%%%%%%%%%%

\section{Conclusion}

Tidal Disruption of stars by supermassive central black holes (SMBH) from dense star clusters is modelled by high-accuracy direct $N$-body simulation. In an extended parameter study we study the dependence of the tidal accretion rate on particle number and the size of the tidal radius. While the tidal radius astrophysically is the radius, at which tidal forces from the SMBH disrupt a star, we treat it here as a free parameter, which is varied by order of magnitude and different from the real one, for a scaling analysis. Our results show that the loss-cone accretion model by \citet{FR1976} and other authors (see Introduction) works well, and we can show that the nearly all stars which are tidally accreted stem from a region around the critical radius, where the time scales of angular momentum diffusion into the loss cone and accretion to the black hole are balanced. Also the majority of accreted stars are originally unbound to the SMBH.

Our simulations show how does the accretion depend on BH's Brownian motion and the density cusp. In our models cusp formation and the decay of the Brownian motion of the SMBH always happen together and depend on each other; the inner edge of the cusp in Fig.\ref{fig_Mdot_1} coincides approximately with the radius to which the motion of our black hole extends in Fig.\ref{fig_Ndot_bhxv}. There is some evidence, that black hole Brownian motion is the more important process here, because in Fig.\ref{fig_Mdot} we can see, TDR curve of model N30 already begins to fall at $T = 0.5 t_{rh}$, while that of model N31 is still rising. From Fig.\ref{fig_Mdot_1} we note that the density cusp is still growing at $T = 0.5 t_{rh}$. At this time, the main difference between these two models is BH mass and hence the amplitude of Brownian motion. For large tidal radii models, TDR of model N20 is close to its maximum value, TDR of model N10, N00 is still rising, BHs in these models are still not massive enough. Another evidence is in Fig.\ref{fig_Mshell-Rmax}, which shows that most of the accreted mass stems from the outer regions of the cusp, near the critical radius. In this figure it is evident that there is some more mass accreted from inside after cusp formation, but in total it is a small fraction. So cusp forming is not the main reason for the fast loss-cone refilling, it only causes the TDR to increase.

Our simulations confirm that in spherically symmetric systems the loss cone for tidal accretion onto SMBH becomes empty after some short initial adjustment time (if the particle number and the tidal radius are large enough). In that regime the refilling of the loss cone is determined by diffusion of angular momentum through two-body relaxation (diffusive regime), which can be long compared to the age of the galaxy or the time until its next merger. We also clearly show that the diffusive regime can only be attained if the SMBH moves only very slowly in the galaxy center; we see that the motion of the SMBH as expected decreases with the mass ratio between the stars and the SMBH (i.e. the particle number in our simulation setup, where the SMBH has always a fixed fraction of the total mass). For the situation of intermediate mass black holes in globular star clusters the mass ratio will not be large enough to realise our diffusive loss cone regime, hence density profiles and kinematics will be different (less cuspy) in such a case. We see from our simulations that in such a case the preference for tidally accreted stars to be on highly eccentric orbits is not so strong, there is a much larger scatter of the eccentricities of the accreted stars.

We do not exactly reproduce the power-law of the Bahcall-Wolf stellar density cusp solution. Both ~\citet{BME2004a} and ~\citet{BBK2011} find similar results in their earlier papers, their power-law as well varying in time and space (between -1.3 and -1.9). There can be different reasons for that, like motion of the black hole, still too low particle numbers in the vicinity of the SMBH or a relatively large tidal radius, since tidal accretion causes a flattening of the density profile \citep{AFS2004}. In our opinion the best numerical reproduction has been found by~\citet{PMS2004}, who did choose carefully an initial model and excluded tidal accretion. In that case they reached -1.75 as power-law coefficient rather precisely. We tried to improve the statistics in our simulations by combining many snapshots into one to obtain a large particle number, but the density center of these snapshots has slightly different positions and it turned out to be hard to improve our current results.

TD also plays a role in destroying the cusp. In \cite{BME2004a}, only about 1\% of the total stars get disrupted because of their choices of tidal radius. While in our simulation a much larger fraction of stars are disrupted, which can further decrease the density in the cusp. \citet{Merritt2010} argued that an enlarged loss-cone may produce a parsec-scale core in galactic center, which is likely the case showing here except that our enlarged loss-cone is an artificial fact. An example is model N30, a density plateau is detected in inner most region of the cluster after 2 $t_{rh}$'s integration (see Figure~\ref{fig_Mdot_3x}).

In real galactic nuclei with SMBH the particle number will always be large enough and the mass ratio between stars and SMBH will be small enough to guarantee that the diffusive regime is realised. An example is the SMBH in our own Milky Way, where the velocity of the central SMBH, if we identify it with the radio source SgrA${}^\star$ in the galactic center, is only of the order (little less) than one kilometer per second \citep{RB2004}. From our scaling study we can extrapolate to the real situation in our own galaxy and predict a tidal star accretion rate of $10^{-5}$ - $10^{-6}$ per year, which is in good agreement with other results, especially \cite{BBK2011}.

For future work we plan to improve our models in two ways. First of all, to have a better support for our scaling model we would like to add more simulations with larger particle number (up to one million or more), to get closer to the ``real'' case in terms of (i) the mass ratio between stars and SMBH and (ii) the size of the tidal radius relative to the gravitational influence radius of the black hole. Increasing the black hole mass would help for (i), but not for (ii). So, the only choice is to increase the particle number to achieve both. The parallel performance of our code will allow this and we can in the future use more of the largest GPU clusters in China and elsewhere. A significantly larger particle number will also be essential for more realism in the population of stars accreted to the black hole, to give the stars a stellar mass spectrum, allow for their evolution, distinguish between main sequence stars, giants and compact remnants (black holes, neutron stars and white dwarfs), all of which will have individually different tidal radii (or may be accreted by the SMBH as a whole).

Second, it is clear that the spherically symmetric case is only a very ideal case - even if the star cluster deep in the potential well of the SMBH is spherical, outside, in the region getting unbound to the SMBH, where most of the loss cone stars originate, the stellar system in the galactic nucleus will most probably be perturbed, by bars or tidal perturbations, or some remaining asymmetries from the last merger event. In such a case we expect strongly increased angular momentum diffusion, possibly full loss cone. This is subject of an ongoing work right now.

%%%%%%%%%%%%%%%%%%%%%%%%%%%%%%%

\acknowledgments
\section*{Acknowledgements}

We acknowledge support by Chinese Academy of Sciences through the Silk
Road Project at NAOC, through the Chinese Academy of Sciences Visiting
Professorship for Senior International Scientists, Grant Number $2009S1-5$
(RS), and through the ``Qianren''special foreign experts program of China.
We acknowledge financial support by National Science Foundation of China
(NSFC) under grant No. 11073025

SZ thanks the ``Global Networks and Mobility Program'' of the University
of Heidelberg (ZUK 49/1 TP14.8 Spurzem) for financial support of a research
visit to Heidelberg.

The special GPU accelerated supercomputer {\tt laohu} at the Center of
Information and Computing at National Astronomical Observatories, Chinese
Academy of Sciences, funded by Ministry of Finance of People's Republic
of China under the grant $ZDYZ2008-2$, has been used for the simulations.
We also used smaller GPU clusters {\tt titan}, {\tt hydra} and {\tt kepler},
funded under the grants I/80041-043 and I/84678/84680 of the Volkswagen
Foundation and grants 823.219-439/30 and /36 of the Ministry of Science,
Research and the Arts of Baden-W\"urttemberg, Germany.

Some code development was also done on the Milky Way supercomputer, funded by
the Deutsche Forschungsgemeinschaft (DFG) through Collaborative Research Center
(SFB 881) ``The Milky Way System'' (subproject Z2), hosted and co-funded by the
J\"ulich Supercomputing Center (JSC).

PB acknowledge the special support by the NAS Ukraine under the Main Astronomical
Observatory GPU/GRID computing cluster project.

We thank the authors of \citet{JYM2012} for providing some of their original data,
which have been used in Fig.~\ref{fig_Mdot_comp}.

%%%%%%%%%%%%%%%%%%%%%%%%%%%%%%%%%%%
% bibliography
%%%%%%%%%%%%%%%%%%%%%%%%%%%%%%%%%%%

\bibliographystyle{apj}
\bibliography{TD-2013-new}

%%%%%%%%%%%%%%%%%%%%%%%%%%%%%%%

\appendix

%%%%%%%%%%%%%%%%%%%%%%%%%%%%%%%

\section{Details for Scaling}

First of all, we define some variables that will be used in following derivation. Define $N_r$ for star numbers in real world and $N_m$ for that in models. Define $r_{t,r}$ for tidal radius in real world and $r_{t,m}$ in models.

i) expanding to larger $N$. As discussed in Section~\ref{TDR}, there are two types of scaling formula depending on whether loss-cone is full. For empty loss-cone case, we have

\begin{equation}
\dot N_{r} = \frac{\ln(\Lambda N_{r})}{\ln(\Lambda N_{m})} \dot N_{m} ~[T]^{-1}
\label{Eq_D01}
\end{equation}

For full loss-cone case, we have

\begin{equation}
\dot N_{r} = \frac{N_{r}} {N_{m}}\dot N_{m} ~[T]^{-1}
\label{Eq_D02}
\end{equation}

ii) shrink of tidal radius. We can calculate the $r_{t,r}$ in physical unit by~\citep{FR1976}

\begin{equation}
r_{t,r} = 1.4 \times 10^{11} (\frac{M_{\bullet}} {M_{\odot}})^{1/3} ~\rm cm
\label{Eq_D03}
\end{equation}

and we have

\begin{equation}
%[L] = \frac{1.09}{\widetilde{r_h}} (\frac{M_{\bullet}}{M_{\odot}})^{0.54} ~\rm pc
[L] = \frac{1.94 \times 10^{15}}{\widetilde{r_h}} (\frac{M_{\bullet}}{M_{\odot}})^{0.54} ~\rm cm
\label{Eq_D04}
\end{equation}

here $\widetilde{r_h}$ is the dimensionless influence radius in $N$-body unit ($r_h = \widetilde r_h ~[L]$). These two equations give out

\begin{equation}
r_{t,r} = 7.2 \times 10^{-5} \widetilde{r_h} (\frac{M_{\bullet}} {M_{\odot}})^{-0.21} ~[L]
\label{Eq_D05}
\end{equation}

From simulation we find that $\dot{N} \propto r_t^{0.39}$, so we have

\begin{equation}
\dot N_r = \dot N_m \times (\frac{r_{t,r}}{r_{t,m}})^{0.39} ~[T]^{-1}
\label{Eq_D06a}
\end{equation}

In the case of empty loss-cone, combine these two scaling process will result in

\begin{equation}
\dot N_r = \dot N_m \times \frac{\ln(\Lambda N_{r})}{\ln(\Lambda N_{m})} \times (\frac{r_{t,r}}{r_{t,m}})^{0.39} ~[T]^{-1}
\label{Eq_D06b}
\end{equation}

iii) up to now all results are expressed in the unit of $[T]^{-1}$, we need to substitute $[T]$ with physical unit, which is done by
following equation

\begin{equation}
[T] = 2.36 \times 10^2 \sqrt{\frac{\widetilde M(\widetilde r_h)}{\widetilde r_h^3} }
      (\frac{M_{\bullet}}{M_{\odot}})^{0.31} ~\rm yr
\label{Eq_D07}
\end{equation}

\noindent where $\widetilde M(\widetilde r_h)$ is the dimensionless enclosed mass within $\widetilde r_h$. And we get a time dependence of $M_{\bullet}$, which read $[T] \propto M_{\bullet}^{0.31}$.

As mentioned above, we assume the mean stellar mass to be 1 solar mass and BH mass is 1\% of the total mass. We can substitute $N_{r}$ with $M_{\bullet}$ using these assumptions and get $N_r = 100 M_{\bullet} / M_{\odot}$. Also substitute $r_{t,r}$ with Eq.~\ref{Eq_D05}, $r_{t,m}$ with $\widetilde r_{t,m}$ and $[T]$ with Eq.~\ref{Eq_D07}. We get

\begin{equation}
\dot N_r = 1.03 \times 10^{-4} \frac{\widetilde r_h^{1.89}}{\widetilde r_{t,m}^{0.39} \widetilde M(\widetilde r_h)^{0.5}}
            \frac{\dot N_m}{\ln(\Lambda N_m)} F(M_{\bullet}) ~\rm yr^{-1}
\label{Eq_D08}
\end{equation}

\noindent where $F(M_{\bullet}) = \ln(100\Lambda M_{\bullet}/M_{\odot})(\frac{M_{\bullet}}{M_{\odot}} )^{-0.39}$. In the mass range we discussed here $F(M_{\bullet})$ is a decreasing function of $M_{\bullet}$, so more massive BH have smaller TDR in empty loss-cone regime.

For full loss-cone regime, following the same procedure one will find

\begin{equation}
\dot N_r = 1.03 \times 10^{-2} \frac{\widetilde r_h^{1.89}}{\widetilde r_{t,m}^{0.39} \widetilde M(\widetilde r_h)^{0.5}}
            \frac{\dot N_m}{N_m} (\frac{M_{\bullet}}{M_{\odot}})^{0.61} ~\rm yr^{-1}
\label{Eq_D09}
\end{equation}

If we choose the following parameters: $\Lambda = 0.11$, $\widetilde r_h = 0.1$, $\widetilde r_{t,m} = 10^{-3}$, $\widetilde M(\widetilde r_h) = 0.01$, $\dot N_m = 11.4$ and $N_m = 128$ K. Eq.~\ref{Eq_D08} can be further converted into

\begin{equation}
\dot N_r = 2.34 \times 10^{-4} F(M_{\bullet}) ~\rm yr^{-1}
\label{Eq_D10}
\end{equation}
\noindent for BH mass between $10^6$ and $10^8 ~M_{\odot}$, $F(M_{\bullet}) \in [1.58\times10^{-2} , 7.41\times10^{-2}]$.

Eq.~\ref{Eq_D09} can be further converted into

\begin{equation}
\dot N_r = 1.71 \times 10^{-6} (\frac{M_{\bullet}}{M_{\odot}})^{0.61} ~\rm yr^{-1}
\label{Eq_D11}
\end{equation}

%%%%%%%%%%%%%%%%%%%%%%%%%%%%%%%%%%%%%%%%%%%%%%%%%%%%%%%%%%%%%%%%%%%%%%%%%%%%%%%%%%%%%%%%%%%%
\end{document}